\documentclass[a4paper,11pt]{article}
\pdfoutput=1 %

\usepackage{jheppub} 
\usepackage{amsmath,amssymb,graphicx,float,slashed,xcolor,multicol}
\usepackage{graphicx,tabularx}
\usepackage{url}
\usepackage{footmisc}
\usepackage{amsfonts}
\usepackage{cancel}
\usepackage{color}
\usepackage{multirow} 
\usepackage{amssymb}
\usepackage{pifont}
\usepackage{epstopdf}
\usepackage{slashed}
\usepackage{comment}
\usepackage{booktabs} 
\usepackage{natbib}
\usepackage{array}
\usepackage{mathrsfs}
\usepackage{subcaption}
\usepackage[toc,page]{appendix}
\usepackage{mathtools}
\usepackage{romannum}
\usepackage{ulem}
\usepackage{bbold}
\usepackage{enumitem}
\usepackage{multirow}

\usepackage{amsmath,amssymb,graphicx,float,slashed,xcolor,multicol}
\usepackage{caption,subcaption}
\usepackage{float}
\usepackage{multirow}
\usepackage{slashed}
\usepackage{upgreek}
\usepackage[toc,page]{appendix}
\usepackage{romannum}
\allowdisplaybreaks

\newcommand*{\rom}[1]{\expandafter\@slowromancap\romannumeral #1@}

\def\lag{\mathscr{L}}

\def\ra{\rightarrow}

\def\beq{\begin{equation}}
\def\eeq{\end{equation}}
\title{Nonstandard neutrino interactions at COHERENT, DUNE, T2HK and LHC}
\author[a]{Tao Han,}
\author[b]{Jiajun Liao,}
\author[a]{Hongkai Liu,}
\author[c]{Danny Marfatia}
\affiliation[a]{Department of Physics and Astronomy, University of Pittsburgh, Pittsburgh, PA 15260, USA }
\affiliation[b]{School of Physics, Sun Yat-Sen University, Guangzhou, 510275, China}
\affiliation[c]{Department of Physics and Astronomy, University of Hawaii at Manoa, Honolulu, HI 96822, USA}
\emailAdd{than@pitt.edu}
\emailAdd{liaojiajun@mail.sysu.edu.cn}
\emailAdd{hol42@pitt.edu}
\emailAdd{dmarf8@hawaii.edu}

\preprint{
\begin{flushright}
PITT-PACC-1815
\end{flushright}
}

\abstract{We study how nonstandard neutrino interactions (NSI) may be probed by a combination of coherent elastic neutrino-nucleus scattering, neutrino oscillation and collider data, from COHERENT, DUNE, T2HK and the high-luminosity (HL) LHC. We focus on NSI induced by a new flavored gauge boson $Z'$ in a generic anomaly-free ultraviolet-complete model. For $Z'$ masses above 10~GeV, the HL-LHC has the best sensitivity regardless of the flavor structure of the model. For masses between 0.01~GeV$-10$~GeV, current LHCb  data and future COHERENT data have the best sensitivity unless the $Z'$ couplings to the first and second generation leptons are suppressed, in which case DUNE and T2HK have the best sensitivity. For $Z'$ masses between about 5~MeV$-20$~MeV, DUNE and T2HK have the best sensitivity.
We also show how joint analyses of COHERENT and LHC data may constrain such models.}

\begin{document}

\titlepage

\maketitle



\flushbottom

\section{Introduction}
\label{sec:intro}

Neutrino oscillations have been confirmed by many neutrino experiments using solar, atmospheric, reactor, and accelerator neutrinos in the last two decades. Since the explanation of neutrino oscillations requires nonvanishing neutrino masses,  the observation of neutrino oscillation provides clear evidence of new physics beyond the Standard Model (SM)~\cite{Tanabashi:2018oca}. The next generation precision neutrino oscillation experiments, DUNE and T2HK,
will have the sensitivity to probe new physics beyond the standard three neutrino paradigm. A model-independent way of studying new physics in neutrino oscillations was first formulated in Ref.~\cite{Wolfenstein:1977ue}, and is now generalized 
in the framework of an effective field theory for nonstandard interactions (NSI);  for reviews  see Ref.~\cite{Ohlsson:2012kf, Miranda:2015dra, Farzan:2017xzy}. In this framework, NSI not only affect neutrino propagation in matter via
neutral current interactions, but also affect neutrino production and
detection via charged current interactions. Since model-independent bounds on the charged current NSI involving charged leptons are generally an order of magnitude stronger than the neutral current NSI~\cite{Biggio:2009nt}, we neglect charged current NSI in this work, and focus on 
neutral current NSI. 

In general, neutral current NSI can be described by dimension-six four-fermion operators of the form~\cite{Wolfenstein:1977ue, Guzzo:1991hi},
\beq
\begin{split}
	\lag_{\rm{NSI}}& =-2\sqrt{2}G_F\sum_{C}\epsilon_{\alpha\beta}^{fP}(\bar{\nu}_\alpha\gamma^\mu P_L \nu_\beta) (\bar{f}\gamma_\mu P_C f)\\
	&=-\sqrt{2}G_F\epsilon_{\alpha\beta}^{fV}(\bar{\nu}_\alpha\gamma^\mu P_L \nu_\beta) (\bar{f}\gamma_\mu f)-\sqrt{2}G_F\epsilon_{\alpha\beta}^{fA}(\bar{\nu}_\alpha\gamma^\mu P_L \nu_\beta) (\bar{f}\gamma_\mu \gamma^5 f)
	\label{eq:NSI}\,,
\end{split}
\eeq
where $\alpha, \beta$ label the lepton flavors ($e, \mu, \tau$), $f$ denotes the fermion fields ($u,d,e$), and $C$ indicates the chirality ($L, R$). Here,
\beq
\epsilon_{\alpha\beta}^{fV} \equiv \epsilon_{\alpha\beta}^{fL} + \epsilon_{\alpha\beta}^{fR}\,, \quad \epsilon_{\alpha\beta}^{fA} \equiv \epsilon_{\alpha\beta}^{fR} - \epsilon_{\alpha\beta}^{fL}\,,
\eeq
with 
$\epsilon^{fL}_{\alpha\beta}$, $\epsilon^{fR}_{\alpha\beta}$ being dimensionless parameters that quantify the
strength of the new interactions in units of the Fermi constant, $G_F\equiv (\sqrt{2}v^2_{\rm{EW}})^{-1}$, with $v_{\rm{EW}}= 246$ GeV, the electroweak scale. These contact interactions arise as a result of integrating out a vector mediator significantly heavier than the typical momentum transfer of the processes. As such, the dimensionless coupling parameters are naturally of the order of $\epsilon \sim g'^2 v^2_{\rm{EW}}/M^2$, where $M$ and $g'$ are  the mediator's mass and coupling.
Similar to the standard matter effect ~\cite{Wolfenstein:1977ue, Mikheev:1986gs}, neutral current NSI affect neutrino propagation in matter via coherent forward scattering, in which the momentum transfer is negligibly small compared with other relevant scales involved. Therefore, the adoption of effective four-fermion interactions in Eq.~(\ref{eq:NSI}) is well justified regardless of the mass of the mediator that induces NSI. Also, for neutrinos propagating in unpolarized matter at rest, only the vector combination contributes to the matter potential.

Coherent elastic neutrino-nucleus scattering (CE$\nu$NS), which was first observed by the COHERENT experiment in 2017 in a cesium iodide (CsI) scintillation detector~\cite{Akimov:2017ade}, provides another sensitive probe of new vector neutral current interactions. CE$\nu$NS occurs when the momentum transfer $Q$ during neutrino scattering off a nucleus is smaller than inverse of the nuclear radius $R$. In the process, the scattering amplitudes of the nucleons inside a nucleus are in phase and add coherently, which leads to a large enhancement of the cross section. In the SM, CE$\nu$NS is induced via the exchange of a $Z$ boson~\cite{Freedman:1973yd}. Hence, CE$\nu$NS is also sensitive to NSI induced by a new neutral vector boson~\cite{Barranco:2005yy,Scholberg:2005qs}. 
%
To probe NSI at higher energies, the formulation of Eq.~(\ref{eq:NSI}) may no longer be valid. 
First, the momentum-dependent propagator of the mediator should be used if its mass $M$ is not much larger than the typical momentum transfer $Q$ to properly model the energy dependence.
Second, SM gauge-invariant operators must be adopted when the momentum transfer or the mediator mass is at or above the electroweak scale.
Thus, an underlying
model that generates NSI is often required.  In fact, most of the new physics scenarios  associated with the lepton sector  at high energies yield NSI~\cite{Antusch:2008tz, Biggio:2009nt, Farzan:2017xzy}. 

\begin{figure}[tb]
	\centering
	\begin{subfigure}{.49\textwidth}
		\centering
		\includegraphics[width=\textwidth]{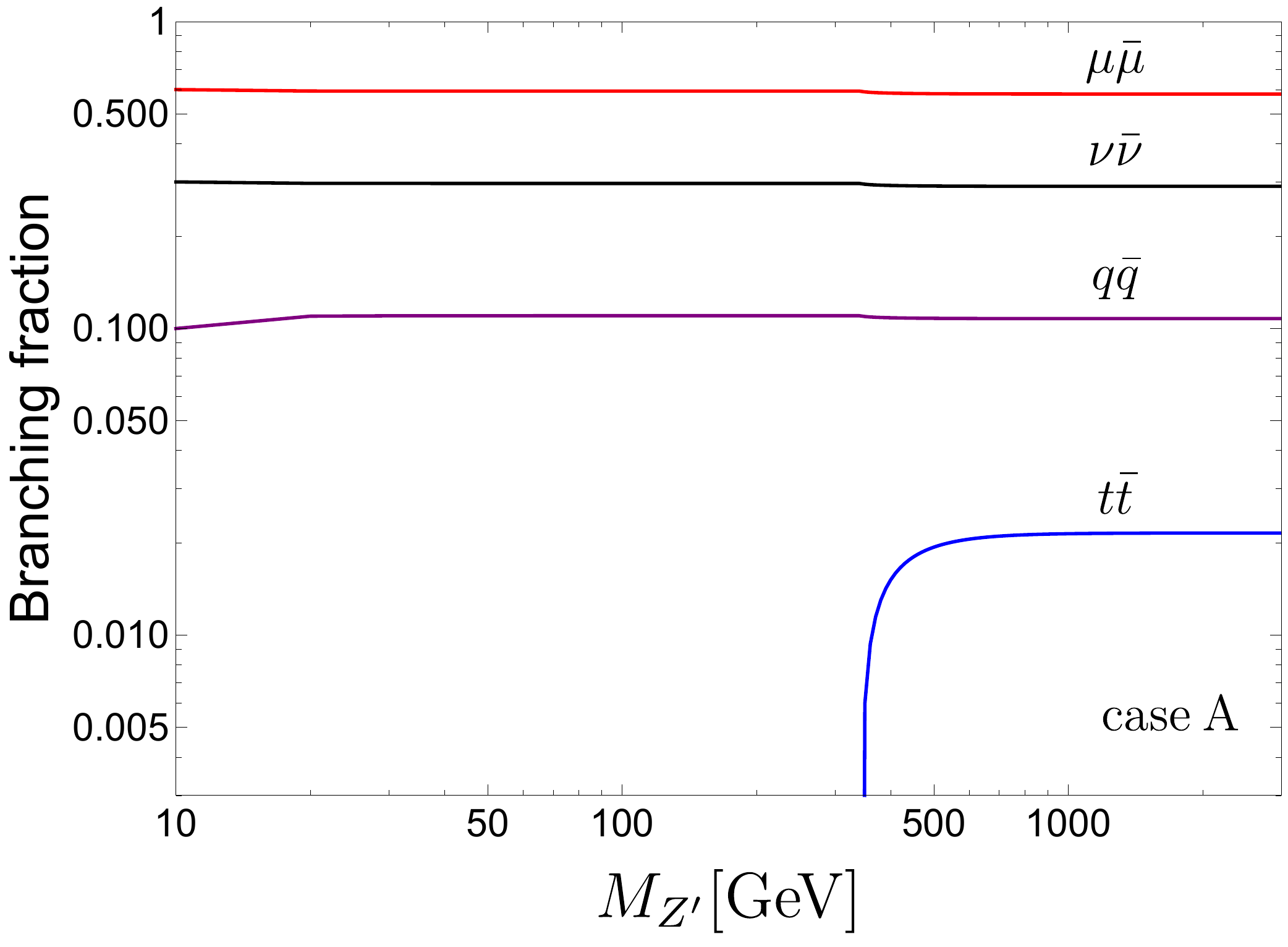}
		\label{fig:decayA}
	\end{subfigure}
	\begin{subfigure}{.49\textwidth}
		\centering
		\includegraphics[width=\textwidth]{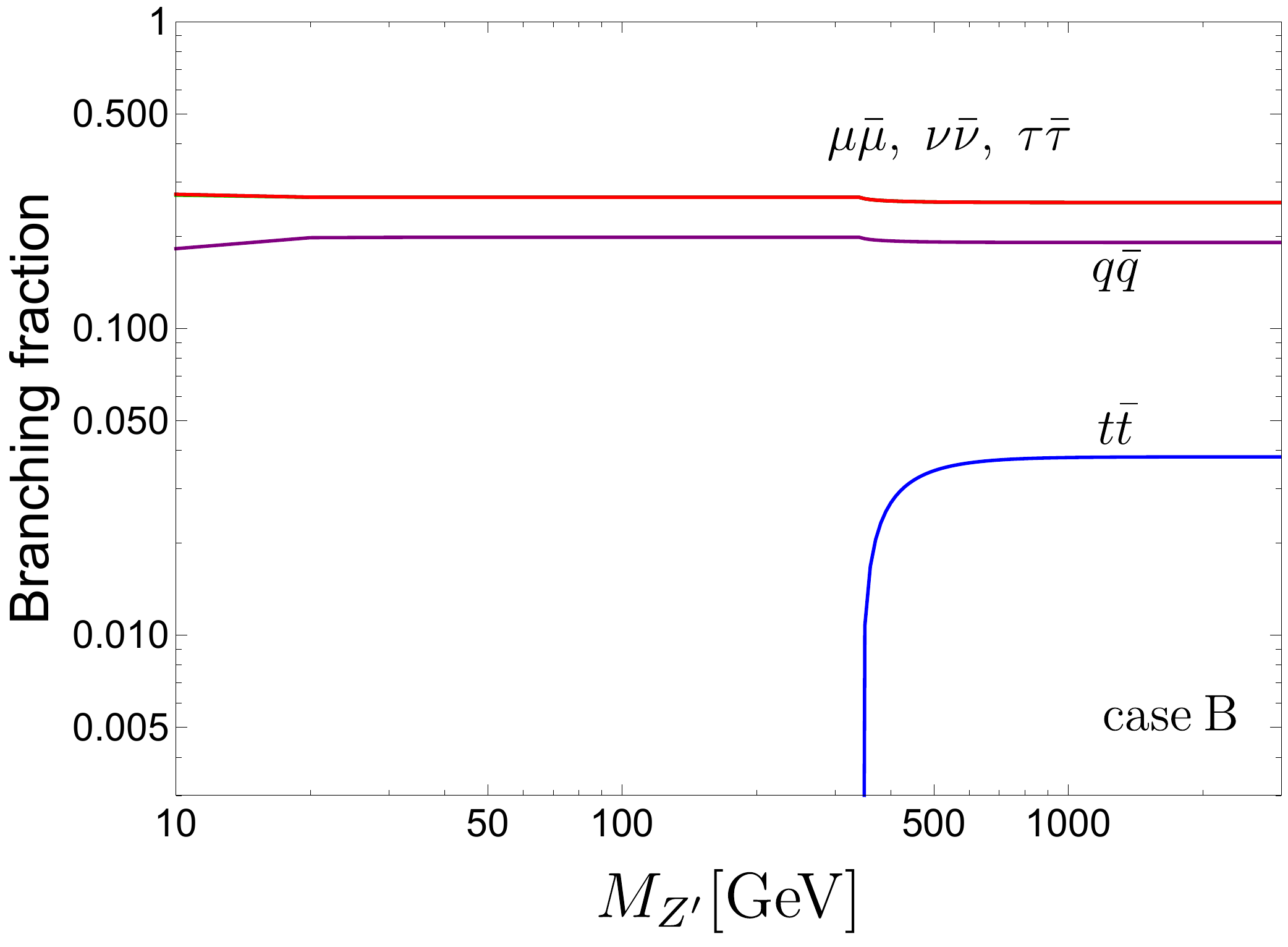}
		\label{fig:decayB}
	\end{subfigure}
	\begin{subfigure}{.49\textwidth}
		\centering
		\includegraphics[width=\textwidth]{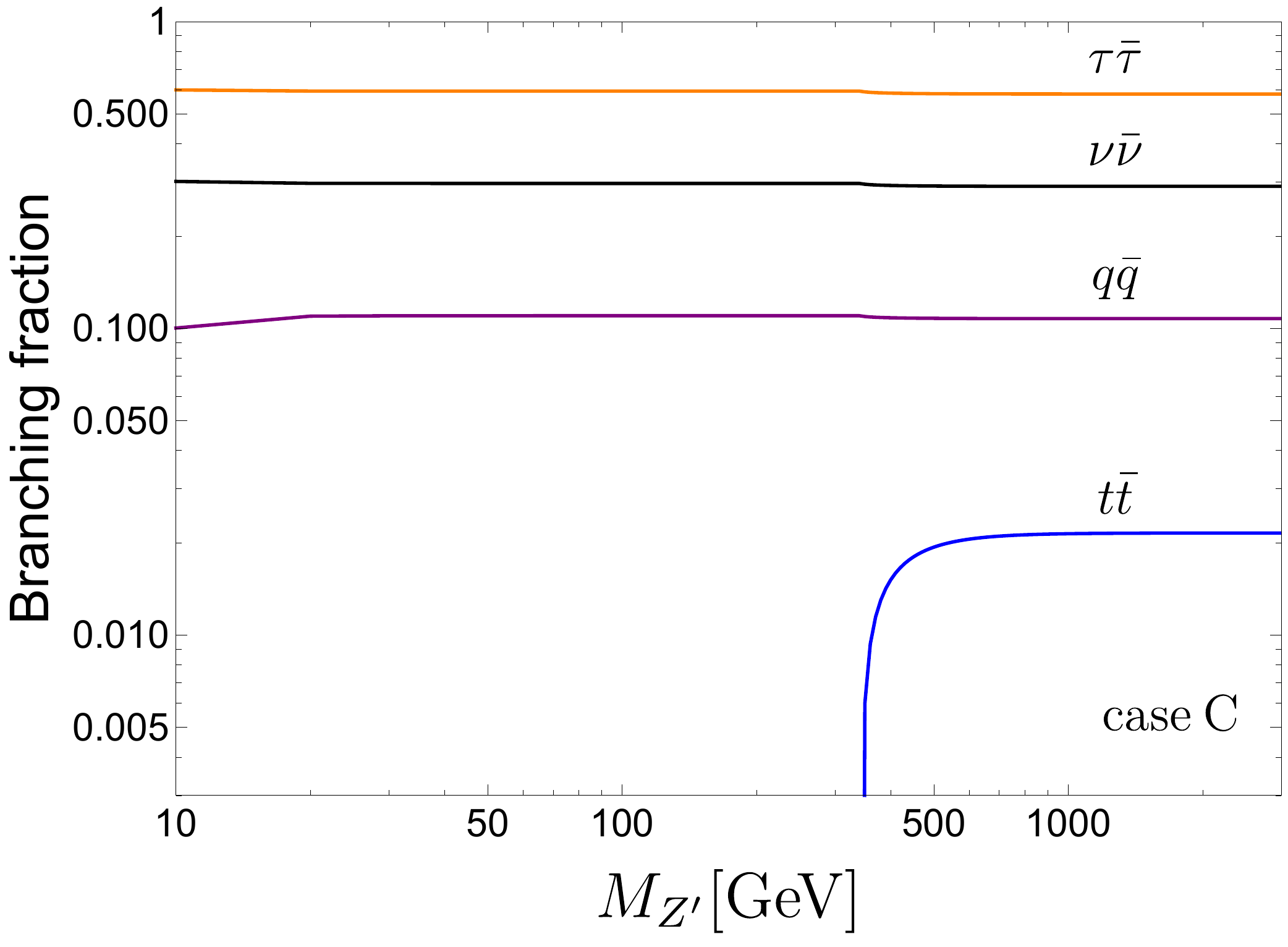}
		\label{fig:decayC}
	\end{subfigure}
	\caption{The branching fractions of $Z^\prime$ for Case A (upper right), B (upper left), and C (bottom), with $q=\{u,d,c,s,b\}$.}
	\label{fig:decay}
\end{figure}

In this paper we focus on a simple model in which the NSI is induced by a gauge boson $Z^\prime$ associated with a new $U(1)^\prime$ symmetry. Assuming the presence of three right-handed neutrinos, the most general anomaly-free $U(1)^\prime$ model can be generated by 
\beq
X = Q_1^\prime B_1 + Q_2^\prime B_2 +Q_3^\prime B_3 + Q_e^\prime L_e+Q_\mu^\prime L_\mu+Q_\tau^\prime L_\tau\,,
\eeq
with the quark charges $Q_{1,2,3}^\prime$ and lepton charges  $Q_{e,\mu,\tau}^\prime$ satisfying the constraint~\cite{Kownacki:2016pmx}
\beq
3(Q_1^\prime+Q_2^\prime+Q_3^\prime)+Q_e^\prime+Q_\mu^\prime+Q_\tau^\prime=0\,.
\eeq
We further require $Q_1^\prime=Q_2^\prime=Q_3^\prime=Q_q^\prime$ to avoid large flavor changing neutral currents in the quark sector. The Lagrangian can be written as
\beq
\lag = \lag_{\rm{SM}} - \frac{1}{4}Z^{\prime \mu \nu} Z^{\prime}_{\mu\nu} + \frac{1}{2}M_{Z^\prime}^2 Z^{\prime \mu} Z^{\prime}_{\mu} + Z^{\prime}_{\mu} J_X^{\mu},
\label{eq:lagZ}
\eeq
where the current\footnote{We have decoupled $\nu_R$ assuming they are heavy and inaccessible.}
\beq
	J_X^\mu = g^{\prime} \left[\sum_{q}Q^{\prime}_q \bar{q}\gamma^\mu q + \sum_{L_\ell=\nu_{\ell L}, \ell} Q^{\prime}_{\ell}\overline{L_\ell}\gamma^\mu L_\ell 
	\right]\,,
\eeq
with $g^{\prime}$ being the $U(1)^\prime$ coupling constant. Since neutrino oscillations are not affected by flavor universal NSI, here we only consider nonuniversal flavor-conserving NSI. Also, because scenarios involving $L_e$ are heavily constrained in the low-mass region by electron beam-dump experiments~\cite{Konaka:1986cb, Riordan:1987aw, Bjorken:1988as, Bross:1989mp, Davier:1989wz, Banerjee:2018vgk}, we set $Q_e^\prime = 0$ and only consider the less constrained eletrophobic NSI. 
For the sake of illustration, we take the following three cases for our benchmark studies~\cite{Heeck:2018nzc}:
\begin{enumerate}[label=(\Alph*)]
	\item $Q_q^\prime = 1/3, Q_\mu^\prime =- 3, Q_e^\prime=Q_\tau^\prime = 0.$
	\item $Q_q^\prime = 1/3, Q_\mu^\prime=Q_\tau^\prime =- 3/2, Q_e^\prime=0.$
	\item $Q_q^\prime = 1/3, Q_\tau^\prime =- 3, Q_e^\prime=Q_\mu^\prime = 0.$
\end{enumerate}
Note that in all  three cases the new gauge boson couples to quarks universally. The partial decay width to a pair of fermions is given by
\beq
\Gamma(Z^\prime\ra f\bar{f})=\frac{N_fQ^{\prime 2}_f g^{\prime 2}}{12\pi M_{Z^\prime}}(M_{Z^\prime}^2+2m_f^2)\sqrt{1-\frac{4m_f^2}{M^2_{Z^\prime}}}\,,
\label{decay}
\eeq
where $N_q=3$, $N_l=1$, and $N_\nu=1/2$. The branching fractions can then be calculated assuming that the total decay width of the $Z^\prime$ is the sum over the SM fermion final states given in Fig.~\ref{fig:decay}. It is important to note that a SM gauge-invariant formulation of NSI often leads to simultaneous couplings to charged leptons due to the symmetry nature of the gauge doublet\footnote{It is possible, though, to arrange for the charged lepton coupling to vanish~\cite{Davidson:2003ha, Ibarra:2004pe}.}  $(\nu, \ell)$.
This opens up new avenues to search for the new physics associated with NSI, 
and it also results in stringent constraints on NSI owing to the correlation with the charged leptons.
As such, the new gauge boson, if heavy,  can be most conveniently searched for at high-energy colliders, especially at the LHC in the di-lepton final state,
\beq
p\,p\rightarrow \ell^+  \ell^- + X\,,
\eeq
where X denoted everything in an inclusive search. For our benchmark choices, we have $\ell = \mu$ for Cases A and B, and $\ell = \tau$ for Case C. We note that in Cases A and B, where muon number $L_\mu$ is involved, one also can make use of $e^+e^- /pp\rightarrow 4\mu$ decays at the $B$-factories and LHC to 
search for a relatively low mass gauge boson. We do not consider $Z'$ bosons lighter than 5~MeV to avoid affecting big bang nucleosynthesis.
%
Once a signal for new physics is observed, it is ultimately important to seek other complementary signals to establish a consistent picture of the underlying physics. 
In this paper we set out to consider correlated signatures between CE$\nu$NS and collider searches.

The rest of the paper is organized as follows. In Section~\ref{sec:Oscillation}, we discuss the current and future sensitivities to NSI from neutrino oscillation experiments. In Section~\ref{sec:CEvNS}, we analyze the current and projected constraints on NSI from the COHERENT experiment.  In Section~\ref{sec:collider}, we study constraints on the model from LHC searches. Correlated studies are presented in Section~\ref{sec:correlation}. We summarize our results in Section~\ref{sec:sum}. 

\section{NSI in neutrino oscillation experiments}
\label{sec:Oscillation}
The Hamiltonian for neutrino propagation in the presence of neutral current NSI is
\begin{equation}
H = {1\over2E}  U
\left( \begin{array}{ccc}
	0 & 0 & 0\\ 0 & \delta m^2_{21} & 0 \\ 0 & 0 & \delta m^2_{31}
\end{array} \right)
U^\dagger + V\,,
\end{equation}
where $U$ is the Pontecorvo-Maki-Nakagawa-Sakata (PMNS) mixing matrix~\cite{Tanabashi:2018oca}
\begin{equation}
U = \left(\begin{array}{ccc}
	c_{13} c_{12} & c_{13} s_{12} & s_{13} e^{-i\delta}
	\\
	-s_{12} c_{23} - c_{12} s_{23} s_{13} e^{i\delta} &
	c_{12} c_{23} - s_{12} s_{23} s_{13} e^{i\delta} &
	c_{13} s_{23}
	\\
	s_{12} s_{23} - c_{12} c_{23} s_{13} e^{i\delta} &
	-c_{12} s_{23} - s_{12} c_{23} s_{13} e^{i\delta} &
	c_{13} c_{23}
\end{array} \right)\,,
\end{equation}
and $V$ is the potential from interactions of neutrinos in matter, which can be expressed using the NSI operators in Eq.~(\ref{eq:NSI}) as
\begin{align}
V =  V_{CC} \left(\begin{array}{ccc}
	1 + \epsilon_{ee} & \epsilon_{e\mu} & \epsilon_{e\tau}
	\\
	\epsilon_{e\mu}^* & \epsilon_{\mu\mu} & \epsilon_{\mu\tau}
	\\
	\epsilon_{e\tau}^*& \epsilon_{\mu\tau}^* & \epsilon_{\tau\tau}
\end{array}\right)\,.
\label{eq:V}
\end{align}
Here,
$V_{CC} \equiv \sqrt2 G_F N_e $, is the standard matter potential, and the effective NSI parameters are
\begin{equation}
\epsilon_{\alpha\beta}\equiv\sum\limits_{q}\epsilon^{qV}_{\alpha\beta}\frac{N_q}{N_e}
\end{equation}
with $N_{q,e}$ the number density of fermions $q=u,d$ and $e$.

Since neutrino propagation in matter is affected by coherent forward scattering, in which the momentum transfer is zero, the effective Lagrangian from Eq.~(\ref{eq:lagZ}) that is relevant for NSI can be written as
\beq
\lag_\text{eff}=-\frac{(g^\prime)^2}{M_{Z^\prime}^2}\left[\sum_{q}Q^{\prime}_q \bar{q}\gamma^\mu q \right]\left[\sum_{\alpha}Q^{\prime}_\alpha \bar{\nu}_\alpha\gamma^\mu P_L\nu_\alpha\right]\,,
\label{eq:Leff}
\eeq
regardless of the $Z^\prime$ mass. Comparing Eqs.~(\ref{eq:NSI}) and~(\ref{eq:Leff}), we have 
\begin{align}
\epsilon^{qV}_{\alpha\alpha} =\frac{(g^\prime)^2Q^\prime_\alpha Q^\prime_q}{\sqrt{2}G_F M_{Z^\prime}^2 }\,.
\label{eq:epsf}
\end{align}
We can then use the bounds on the NSI parameters from neutrino oscillation experiments to constrain the parameter spaces in the $Z^\prime$ models. For Case A (C), the model predicts that only $\epsilon_{\mu\mu}$ ($\epsilon_{\tau\tau}$) is nonzero. For Case B,  since $\epsilon_{\mu\mu}$ is equal to $\epsilon_{\tau\tau}$, and neutrino oscillation probabilities are not affected by a subtraction of a diagonal contribution from the full Hamiltonian, we can obtain constraints on Case B from bounds on NSI with only $\epsilon_{ee}$ being nonzero.

\begin{table}
	\begin{center}
		\begin{tabular}{|l|c|c|}
			\hline
			 & Current data & DUNE+T2HK \\
			\hline
			$\epsilon_{ee}^u$ & $[-1.192, -0.802]\oplus[-0.020, +0.456]$ & $[-0.407,-0.270]\oplus[-0.072, +0.064]$\\
			\hline
			$\epsilon_{\mu\mu}^u$ & $[-0.130, 0.152]$ & $[-0.019, +0.018]$\\
			\hline 
			$\epsilon_{\tau\tau}^u$ & $[-0.152, 0.130]$ & $[-0.017, +0.017]$\\
			\hline 
		\end{tabular}
	\end{center}
	\caption{$2\sigma$ allowed ranges for the diagonal NSI parameters from the global analysis of current neutrino oscillation data~\cite{Esteban:2018ppq}, and from a simulation of DUNE and T2HK.
	\label{tab:NSI}}
\end{table}

\begin{figure}[h!]
	\centering
	\begin{subfigure}{.49\textwidth}
		\centering
		\includegraphics[width=\textwidth]{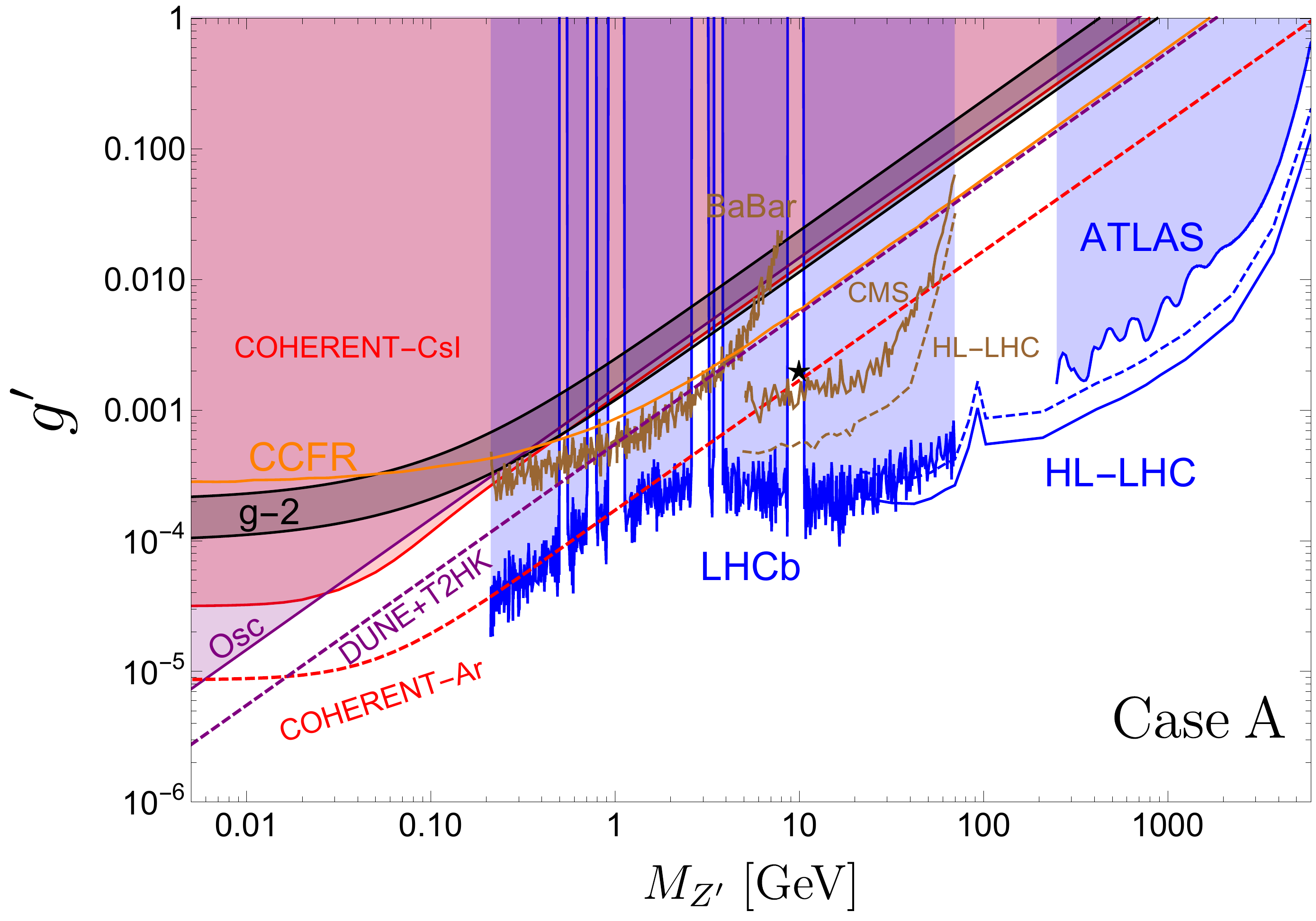}
		\label{fig:caseA}
	\end{subfigure}
	\begin{subfigure}{.49\textwidth}
		\centering
		\includegraphics[width=\textwidth]{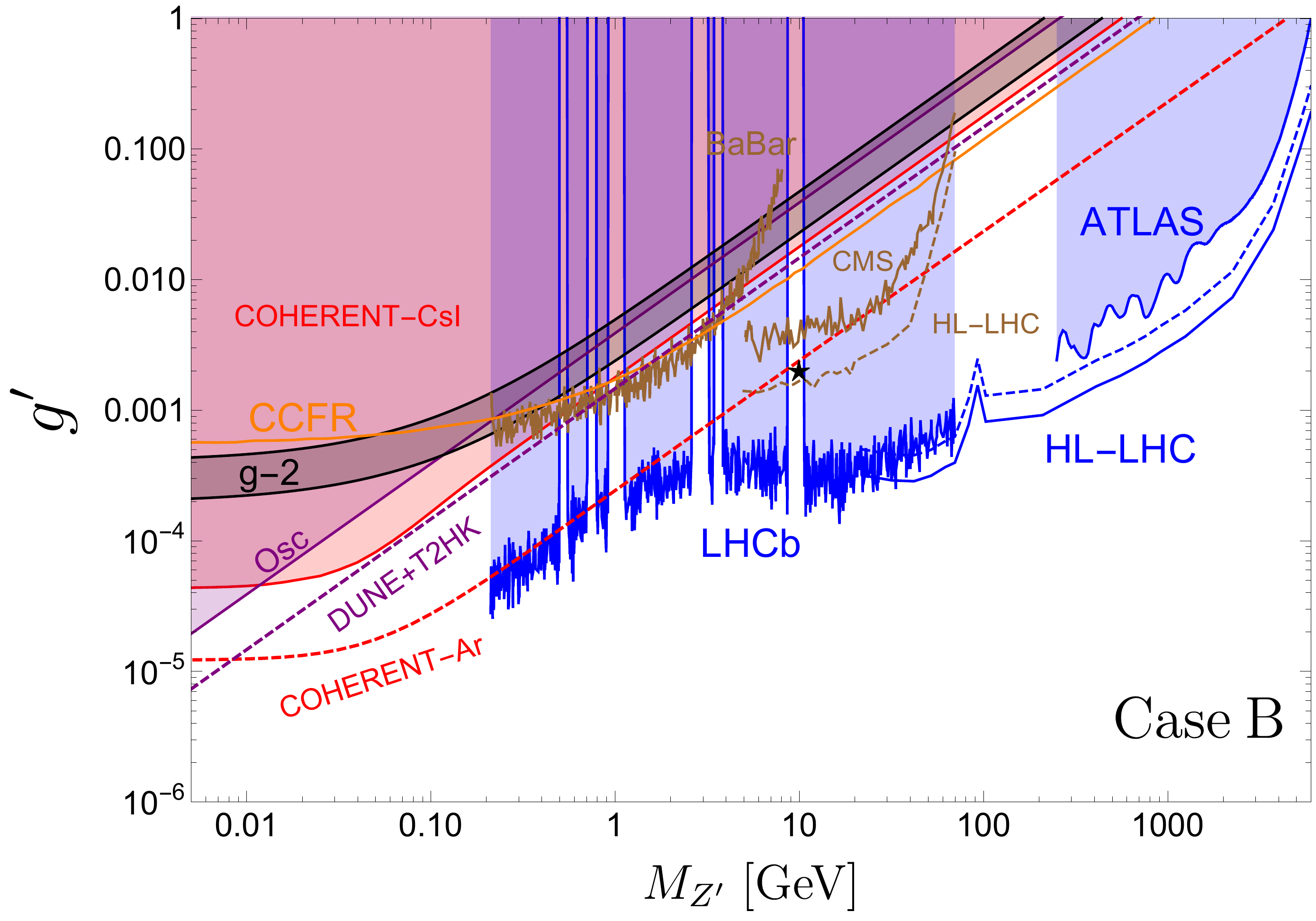}
		\label{fig:caseB}
	\end{subfigure}
	\begin{subfigure}{.49\textwidth}
		\centering
		\includegraphics[width=\textwidth]{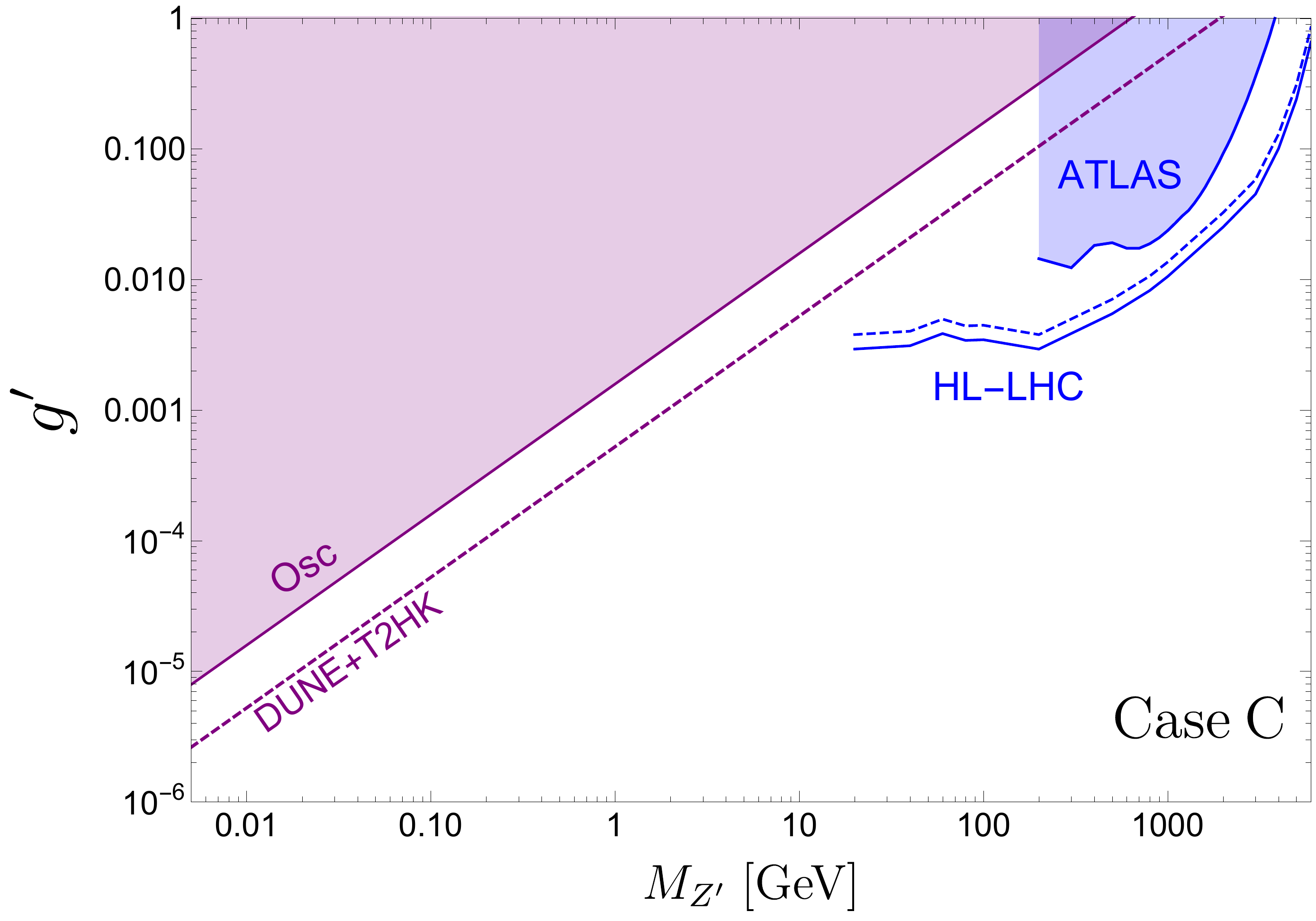}
		\label{fig:caseC}
	\end{subfigure}
	\caption{Bounds on $g^{\prime}$ for Cases A (upper left panel), B (upper right panel) and C~(lower panel). The red shaded areas correspond to the $2\sigma$ exclusion regions by using the energy spectrum from the COHERENT CsI detector~\cite{Akimov:2017ade}. The red dashed lines show the expected 2$\sigma$ limit from COHERENT with a 750~kg LAr detector~\cite{Akimov:2018ghi} and a 4-year exposure using both energy and time information. The purple areas correspond to the $2\sigma$ bounds from a global fit to neutrino oscillation data~\cite{Esteban:2018ppq}. The dashed purple lines show the expected $2\sigma$ exclusion limit from DUNE and T2HK combined. Regions above the brown curves are excluded by CMS~\cite{Sirunyan:2018nnz} and BaBar~\cite{TheBABAR:2016rlg} at $2\sigma$ and 90\% CL, respectively, using $pp/e^+e^- \rightarrow \mu^+\mu^-Z^\prime$ searches. The brown dashed curves are the $2\sigma$ expected sensitivities from HL-LHC, with an integrated luminosity of 3000~$\rm{fb}^{-1}$, in the $\mu^+\mu^-Z^\prime$ channel,  and the blue solid (dashed) curves correspond to the expected 2$\sigma$ (5$\sigma$) limit using di-muon searches for Cases A and B, and di-tau searches for Case C. In the upper panels, the blue shaded regions are excluded at 90\% CL by the LHCb dark photon searches~\cite{Aaij:2019bvg} and at $2\sigma$ by the ATLAS di-muon searches~\cite{Aad:2019fac} with 139 $\rm{fb}^{-1}$. In the lower panel, the blue area is excluded at $2\sigma$ by the ATLAS di-tau searches~\cite{Aaboud:2017sjh} with 36.1 $\rm{fb}^{-1}$. The 2$\sigma$ limit from CCFR~\cite{Mishra:1991bv,Altmannshofer:2014pba} is given by the orange curves. The 2$\sigma$ allowed regions that explain the discrepancy in the anomalous magnetic moment of the muon ($\Delta a_\mu = (29\pm 9 )\times 10^{-10}$~\cite{Jegerlehner:2009ry}) are indicated by the black band. 
	The black stars mark the benchmark points we consider in Section~\ref{sec:correlation}.
	}
	\label{fig:bounds}
\end{figure}

We adopt the 2$\sigma$ bounds on $\epsilon_{\alpha\alpha}^u$ from the global analysis of current oscillation data~\cite{Esteban:2018ppq} as compiled in Table~\ref{tab:NSI}. Note that neutrino oscillation data constrain differences between two diagonal $\epsilon$'s, not individual diagonal $\epsilon$'s. To obtain bounds on a single $\epsilon$, we set one of the two $\epsilon$'s to be zero. We bound $\epsilon_{\mu\mu}^u$ by choosing the smaller of the values obtained by setting $\epsilon_{ee}^u=0$ in $\epsilon_{ee}^u-\epsilon_{\mu\mu}^u$ and $\epsilon_{\tau\tau}^u=0$  in 
 $\epsilon_{\tau\tau}^u-\epsilon_{\mu\mu}^u$. We apply them to constrain the theory parameter space in the $(M_{Z^\prime}, g^\prime)$ plane and the exclusion regions are shown as the purple areas in Fig.~\ref{fig:bounds}. 
Note that the bounds from the global analysis are obtained under the assumption that all NSI parameters are nonzero and then projected to one NSI parameter. Since degeneracies among NSI parameters can significantly weaken the constraints on an individual NSI parameter~\cite{Liao:2016hsa}, the current bounds from the global analysis of oscillation data should be considered to be conservative. 

We also consider the sensitivity of the next generation long-baseline neutrino oscillation experiments, DUNE~\cite{DUNE} and T2HK~\cite{T2HK}. We follow the procedure of Ref.~\cite{Liao:2016orc}, and simulate the DUNE and T2HK data assuming the normal neutrino mass hierarchy, the neutrino CP phase $\delta=0$, and $\epsilon_{\alpha\alpha}=0$. We scan over both the mass hierarchies, the neutrino oscillation parameters and take only one diagonal $\epsilon_{\alpha\alpha}$ to be nonzero at a time.  The 2$\sigma$ allowed ranges for the diagonal NSI parameters are provided in the last column of Table~\ref{tab:NSI}. The expected sensitivities in the $(M_{Z^\prime}, g^\prime)$ parameter space are shown as the purple dashed lines in Fig.~\ref{fig:bounds}. As expected, it simply scales linearly with $g' / M_{Z'}$. The reaches for the three cases are roughly similar. For instance, at $M_{Z'}\sim 10$ GeV, the sensitivity for the couplings can reach $g' \sim 0.008 \,(0.02) \, [0.008]$ for Case~A~(B)~[C]. We see that future bounds on NSI will be improved by a factor of a few compared to current bounds, and the current constraints on the parameter space in Case C for $M_{Z^\prime} \lesssim 200$~GeV only come from neutrino oscillation data.

\section{CE$\nu$NS}
\label{sec:CEvNS}
CE$\nu$NS has recently been measured by the COHERENT experiment, which detects neutrinos from the Spallation Neutron Source (SNS) at Oak Ridge National Laboratory. Neutrinos at the SNS \cite{SNS} consist of a prompt component of monoenergetic $\nu_\mu$ from the stopped pion decays, $\pi^+\to \mu^++\nu_\mu$, and two delayed components of $\bar{\nu}_\mu$ and $\nu_e$ from the subsequent muon decays, $\mu^+\to e^++\bar{\nu}_\mu+\nu_e$.  The fluxes of the three neutrino flavors  ($\nu_\mu$, $\bar\nu_\mu$ and $\nu_e$) are well known and given by
\begin{align}
\label{eq:nu-spectra.COHERENT}
\phi_{\nu_\mu}(E_{\nu_\mu})&={\cal{N}}
\frac{2m_\pi}{m_\pi^2-m_\mu^2}\,
\delta\left(
1-\frac{2E_{\nu_\mu}m_\pi}{m_\pi^2-m_\mu^2}
\right) \ ,
\nonumber\\
\phi_{\nu_e}(E_{\nu_e})&={\cal{N}}\frac{192}{m_\mu}
\left(\frac{E_{\nu_e}}{m_\mu}\right)^2
\left(\frac{1}{2}-\frac{E_{\nu_e}}{m_\mu}\right)\ ,
\nonumber\\
\phi_{\bar\nu_\mu}(E_{\bar\nu_\mu})&={\cal{N}}\frac{64}{m_\mu}
\left(\frac{E_{\bar\nu_\mu}}{m_\mu}\right)^2
\left(\frac{3}{4}-\frac{E_{\bar\nu_\mu}}{m_\mu}\right)\,,
\end{align}
where the normalization factor ${\cal{N}}=\frac{rTN_\text{POT}}{4\pi L^2}$, with $r=0.08$ the number of neutrinos per flavor produced per proton collision, $N_\text{POT}=2.1\times 10^{23}$ the total number of protons on target per year, $T$ the number of years of data collection, and $L$ the distance between the source and the detector~\cite{Akimov:2017ade}. The $\nu_\mu$ component has energy $(m^2_{\pi}-m_\mu^2)/(2 m_\pi)\approx 30$~MeV, and the energies of the $\nu_e$ and $\bar\nu_\mu$ have a kinematic upper bound, $m_\mu/2 \approx 50$~MeV. 
The expected number of events with recoil energy in the energy range 
[$E_r$, $E_r + \Delta E_r$] and arrival time in the time interval [$t$, $t+\Delta t$] is given by
\begin{equation}
\label{eq:recoil-spectrum}
N_{th} (t,E_r,\epsilon)=\sum_{\alpha}\frac{m_\text{det}N_A}{M}\int_{\Delta E_r}
\,dE_r\int_{\Delta t}\,dt \rho_\alpha(t) \int_{E_\nu^\text{min}}^{E_\nu^\text{max}}\,dE_\nu\,\phi_\alpha(E_\nu)
\,\frac{d\sigma_{\alpha}(\epsilon)}{dE_r}\, ,
\end{equation}
where $m_\text{det}$ is the detector mass, $M$ is the molar mass
of the target nucleus, $N_A=6.022\times 10^{23}\,\text{mol}^{-1}$, $\rho_\alpha(t)$ is the arrival time Probability Density Function (PDF) provided in the COHERENT data release~\cite{Akimov:2018vzs}, and $\alpha=\nu_\mu,\bar\nu_\mu, \nu_e$. We assume that the presence of new neutral current interactions do not modify the arrival time PDF. 

Neglecting radiative corrections, the differential cross section for a given neutrino flavor $\nu_\alpha$ scattering off a nucleus is given by
\beq
\frac{d\sigma_\alpha(\epsilon)}{dE_r}=\frac{G_F^2}{2\pi}Q_\alpha^2F^2(Q^2)M(2-\frac{ME_r}{E_\nu^2})\,,
\eeq
where $F(Q^2)$ refers to the nuclear form factor taken from Ref.~\cite{Klein:1999gv}. In the presence of NSI, the effective charge can be written as
\beq
Q^2_{\alpha} = [Z (g_p^V+2\epsilon^{uV}_{\alpha\alpha}+\epsilon^{dV}_{\alpha\alpha}) + N (g_n^V+\epsilon^{uV}_{\alpha\alpha}+2\epsilon^{dV}_{\alpha\alpha})]^2\,,
\eeq
where $Z$ ($N$) is the number of protons (neutrons) in the nucleus,  $g_p^V=\frac{1}{2}-2\sin^2\theta_W$ and $g_n^V=-\frac{1}{2}$ are the SM weak couplings, and $\theta_W$ is the weak mixing angle. The NSI parameters for coupling to up and down quarks can be written as
\beq
\begin{split}
	&\epsilon_{ee}^{uV}=\epsilon_{ee}^{dV}=\frac{g^{\prime 2}Q^\prime_qQ^\prime_e}{\sqrt{2}G_F(2ME_r+M_{Z^\prime}^2)}\,,
	\\
	&\epsilon^{uV}_{\mu\mu} = \epsilon^{dV}_{\mu\mu} =\frac{g^{\prime 2}Q^\prime_qQ^\prime_\mu}{\sqrt{2}G_F(2ME_r+M_{Z^\prime}^2)}\,.
\end{split}
\eeq
 For the CsI detector, the total cross section is a sum of the contributions of $^{133}$Cs and $^{127}$I, i.e.,
\begin{equation}
\frac{d\sigma_{\alpha,\text{CsI}}}{dE_r}=\frac{d\sigma_{\alpha,\text{Cs}}}{dE_r}+\frac{d\sigma_{\alpha,\text{I}}}{dE_r}\,.
\end{equation}
To compare with COHERENT data, we convert the nuclear recoil energy to the number of photoelectrons ($n_\text{PE}$) using the relation~\cite{Akimov:2017ade},
\beq
n_\text{PE}=1.17(E_r/\text{keV})\,.
\eeq
Note that we do not use the new quenching factor reported in Ref.~\cite{Collar:2019ihs} as it is still under investigation by the COHERENT collaboration~\cite{Konovalov}. We employ the acceptance function~\cite{Akimov:2018vzs},
\begin{equation}
\label{eq:acceptance}
\mathcal{A}(n_\text{PE})=\frac{k_1}{1+e^{-k_2(n_\text{PE}-x_0)}}\theta(n_\text{PE}-5)\,,
\end{equation}
where $k_1=0.6655$, $k_2=0.4942$, $x_0=10.8507$ and $\theta(x)$ is the Heaviside step function. 

Because the number of events is small and experimental uncertainties large, we use the energy spectrum (but not the timing information) measured by the CsI detector to  evaluate the statistical significance of a nonstandard scenario. We define 
\begin{align}
\chi^2 = \sum_{i=4}^{15} \left[\frac{N_\text{meas}^i-N_\text{th}^i(1+\gamma)-B_\text{on}(1+\beta)}{\sigma_\text{stat}^i}\right]^2+\left(\frac{\gamma}{\sigma_\gamma}\right)^2+\left(\frac{\beta}{\sigma_\beta}\right)^2\,,
\end{align}
where $N_\text{meas}^i$ and $N_\text{th}^i$is the number of measured and predicted events per energy bin, respectively. The statistical uncertainty per energy bin is $\sigma_\text{stat}^i=\sqrt{N_\text{exp}^i+2B_\text{SS}^i+B_\text{on}^i}$, where $B_\text{SS}$ and $B_\text{on}$ are the estimated steady-state and beam-on backgrounds, respectively. $B_\text{SS}$ is determined by the anti-coincident (AC) data, and $B_\text{on}$ mainly consists of prompt neutrons. Both the spectral and temporal distributions of the backgrounds are provided by the COHERENT collaboration~\cite{Akimov:2018vzs}. For the signal normalization uncertainty, we follow the original COHERENT analysis and choose 
$\sigma_\gamma=0.28$, which includes the neutrino flux uncertainty (10\%), form factor uncertainty (5\%), signal acceptance uncertainty 
(5\%), and quenching factor uncertainty (25\%). For the beam-on background uncertainty, we fix $\sigma_\beta=0.25$~\cite{Akimov:2017ade}.
We scan over values of the coupling $g^\prime$ and the mediator mass $M_{Z^\prime}$. The $2\sigma$ exclusion regions in the $(M_{Z^\prime}, g^\prime)$ plane are shown as the red regions in Fig.~\ref{fig:bounds} for Cases A and B. For $M_{Z^\prime}\gtrsim 50$~MeV, the current constraint from COHERENT CsI is comparable to the expected sensitivity of DUNE$+$T2HK for Case B, and is weaker by about a factor of two for Case A. For very small $M_{Z^\prime}$ DUNE+T2HK has greater sensitivty than the current COHERENT bounds for  both Cases A and B. Note that COHERENT data does not place bounds on Case C because the SNS beam does not have $\nu_\tau$ and $\bar\nu_\tau$.

The COHERENT collaboration has an extensive upgrade plan~\cite{Akimov:2018ghi}, part of which is a 750~kg LAr detector located at $L = 29$~m from the source.  We assume a 4-year exposure with the same neutrino production rate as the current setup, which corresponds to 8.4$\times 10^{23}$ protons-on-target (POT) in total. 
Since both the spectral and temporal distributions of the recoil energy events depend on the flavor structure, we perform a two dimensional analysis that utilizes both the spectral and temporal information.
To estimate the projected sensitivities at the LAr detector, we adopt the likelihood function from Ref.~\cite{Dutta:2019eml}, i.e.,
\begin{align}
\mathcal{L}(\vec \theta )\propto& \prod_{(t, E_r)}   \int \int \exp\{ -\lambda(t, E_r)\}\frac{ \{\lambda(t, E_r)\}^{N_{obs}(t, E_r)}}{  N_{obs}(t, E_r)! }
\times 
\frac{\exp (-\gamma^2/2\sigma^2_\gamma)}{\sqrt{\sigma^2_\gamma}}\nonumber \\
&\, \, \, \, \, \, \, \, \times \exp \{- \beta N_{obs,bg}(t, E_r)\}\frac{\{\beta N_{obs,bg}(t, E_r)\}^{N_{obs, bg}(t, E_r)}  }{  N_{obs, bg}(t, E_r)! } \,  \mathrm{d}\gamma \, \mathrm{d}\beta\,.
\label{eq:L}
\end{align}
where $\lambda(t, E_r)= (1+\gamma) N_{th}(t, E_r, \epsilon)+\beta N_{obs,bg}(t, E_r)$. We calculate the number of events expected in the SM for each bin within the range $0<t<6\,\mu s$ and 20~keV $ < E_{\rm{r}}<100$~keV, with bin sizes of 0.5 $\mu s$ and 2 keV, respectively.
We assume that the steady-state background is uniform in energy and is 1/4 of the SM expectation. We also assume the systematic uncertainty 
$\sigma_\gamma$ to be 17.5\%, which corresponds to a reduced quenching factor uncertainty of 12.5\% for LAr. A more precise treatment 
would include energy-dependent form factor uncertainties~\cite{AristizabalSierra:2019zmy}. The projected sensitivities are shown by the purple dashed line in Fig.~\ref{fig:bounds}. A factor of three improvement is expected in the sensitivity to the coupling, compared to the current CsI results.  
We see that future CE$\nu$NS experiments will set stronger bounds than next generation neutrino oscillation experiments for most $Z'$ masses in Cases A and B, and will provide the strongest constraints for 20~$(10) \text{ MeV} \lesssim M_{Z^\prime} \lesssim 1$~GeV in Case A (B). 

\section{Collider searches for NSI}
\label{sec:collider}

As emphasized in the introduction, a SM gauge-invariant formulation of NSI often results in simultaneous couplings to charged leptons. This opens up new avenues to search for the new physics associated with NSI, in particular at colliders. We explore the sensitivity reach at the LHC for NSI via a di-lepton final state from the Drell-Yan (DY) production of a $Z'$,
\begin{eqnarray}
\label{eq:2l}
pp\rightarrow Z^\prime \rightarrow \ell^+\ell^- +X\,,
\end{eqnarray}
with $\ell=\mu, \tau$ and $X$ denotes other inclusive states (like a jet) when kinematically favorable for the signal identification. This is a particularly sensitive signal $M_{Z'} > M_Z$.
We also include a four-lepton final state,
\begin{eqnarray}
\label{eq:4l}
pp\rightarrow Z^*/\gamma^* \rightarrow \ell^+\ell^- + Z' \to \ell^+\ell^- + \ell^+\ell^- +X.
\end{eqnarray}
This channel is more suitable for a low mass $Z'$ as we will see below.

We use the Monte Carlo event generator $\rm{MadGraph5\_aMC@NLO}$~\cite{Alwall:2014hca} to generate signal and background samples with the $\rm{NN23LO1}$ $\rm{PDF}$ set~\cite{Ball:2013hta}. The NSI Lagrangian is implemented in the FeynRules 2.0~\cite{Alloul:2013bka} framework. Pythia 8.1~\cite{Sjostrand:2006za, Sjostrand:2007gs} is used for parton showering and hadronization. Matching is performed with the MLM prescription~\cite{Alwall:2007fs}.  The generated events are passed into $\rm{Delphes}$ $\rm{3.4.1}$~\cite{deFavereau:2013fsa} for fast detector simulation.

\subsection{Cases A and B:  $\mu$ final states}
\label{subsec:dimuon} 
In Case A, the new gauge boson couples to quarks universally,  and only to second generation leptons. While in Case B, the new gauge boson couples equally to second and third generations leptons. We first apply the existing LHC bound on searches for the di-muon final state to both cases, given that muons are much easier to identify than taus at the LHC.  ATLAS~\cite{Aad:2019fac} has performed a search for di-lepton resonances in the 250 ${\rm GeV} \lesssim M_{Z^\prime} \lesssim$ 6~TeV mass range setting a $2\sigma$ upper limit on the fiducial cross section times branching ratio with 139~$\rm{fb}^{-1}$ at $\sqrt{s}=13$~TeV. The fiducial region is defined by the acceptance cuts,
\beq
p^{\mu}_T > 30\ {\rm GeV} ,\quad  | \eta_\mu | < 2.5,\quad 
m_{\ell\ell} > M_{Z^\prime} - 2 \Gamma_{Z^\prime}\,.
\eeq
To extract limits on $g^\prime$, we calculate $\sigma(pp\rightarrow Z^\prime + X)\cdot B(Z^\prime \rightarrow \mu^+\mu^-)$ in the fiducial region at leading order (LO). The expected signal yields are rescaled to next-to-leading order (NLO) accuracy using a K-factor of 1.3~\cite{Khachatryan:2016zqb}. From the auxiliary figure 2c of Ref.~\cite{Aad:2019fac}, the upper limits at $2\sigma$ on the fiducial cross section from ATLAS are translated into the bounds on our model parameters, shown as the blue shaded regions in the upper panels of Fig.~\ref{fig:bounds}.  This search excludes $g^\prime\gtrsim 1.6\; (2.4) \times 10^{-3}$ for $M_{Z^\prime} \approx 250$ GeV in Case A (B). 

Searches for dark photons decaying to di-leptons can shed light on new vector bosons, especially relatively light ones. In Cases A and B, we recast prompt-like dark photon searches at LHCb~\cite{Aaij:2019bvg} to obtain constraints in the mass range 200~MeV to 70~GeV based on the framework developed in Ref.~\cite{Ilten:2018crw}. This is the most sensitive probe currently in this mass window except near the resonances like $J/\psi,\ \Upsilon$ and approaching the $Z$-pole. The corresponding upper limits on the coupling at 90\% CL are shown by the blue shaded regions in Fig.~\ref{fig:bounds}. 

Having discussed the bounds from the di-muon final state, we turn to the four-muon final state.  Both the BaBar and CMS have performed searches for the decay, $\gamma^*/Z^* \rightarrow \mu^+\mu^- Z' \to 4 \mu$. The BaBar searches~\cite{TheBABAR:2016rlg} set a 90\% CL upper limit on the new gauge coupling based on a $L_\mu-L_\tau$ model corresponding to $Q_q^\prime = Q_e^\prime = 0,\ Q_\mu^\prime = -Q_\tau^\prime = 1$ in our parameterization. The CMS searches~\cite{Sirunyan:2018nnz} set a $2\sigma$ upper limit on $g^\prime$  by assuming the branching ratio $B(Z^\prime\rightarrow\mu^+ \mu^-)=1/3$ and $Q^\prime_{\mu}=1$. By rescaling the observed bounds according to the branching fractions and 
production cross section, we extract bounds for our scenarios. The brown curves show the BaBar and CMS bounds in the upper panels of Fig.~\ref{fig:bounds}. We see that the current bound from the LHCb dark photon search is dominant in the medium mass range and disfavors $g^\prime \gtrsim 10^{-4}$ for $M_{Z^\prime} \approx 200$~MeV.

We further estimate the sensitivity reach via the di-muon channel $Z' \to \mu^+\mu^-$ for $10 \lesssim M_{Z^\prime} \lesssim 6000$ GeV at the high-luminosity LHC (HL-LHC) with the full 3000 $\rm{fb}^{-1}$ integrated luminosity. The signal is from the DY process as in Eq.~(\ref{eq:2l}). 
We select events that contain at least two opposite-sign muons. The leading (subleading) muon is required to have $p_T > 22$ (10) GeV. All muons are required to have $|\eta| <$ 2.4. Finally,  in calculating the sensitivity, we apply a mass window cut $0.97\;M_{Z'} < M(\ell^+\ell^-) < 1.03\;M_{Z'}$ below 3 TeV, and use a $3-6$~TeV mass window to ensure enough background events in the high mass region, to optimize the signal observability. The dominant background is from the SM DY process. We also include smaller background contributions from $t\bar{t}$, $tW$, $WW$ and $ZZ$. We generate the signal and DY background with up to two additional jets in the phase space $M_{\mu\mu} < 60$~GeV.  This is so that for a lighter $Z'$, the additional jets help to kick the leptons to a high momentum for more efficient triggering. For $M_{\mu\mu} > 60$ GeV, we generate the signal and DY background at LO and apply the combined QCD and electroweak corrections to the invariant mass distributions according to Ref.~\cite{Barze:2013fru}. $t\bar{t}$ and $tW$ backgrounds are generated at LO and normalized to NNLO + NNLL by a K-factor of 1.84~\cite{web} and 1.35~\cite{Kidonakis:2010ux} respectively. The $WW, WZ,$ and $ZZ$ backgrounds are normalized to NNLO QCD by a K-factor of 1.98~\cite{Gehrmann:2014fva}, 2.07~\cite{Grazzini:2016swo}, and 1.74~\cite{Cascioli:2014yka} respectively. The local significance is defined as
\beq
S_l=\frac{N_{S}}{\sqrt{N_{B}}},
\eeq
where $N_{S}\ (N_B)$ is the expected number of signal (SM background) events. The blue solid (dashed) curves in the upper panels of  Fig.~\ref{fig:bounds} show the $2\sigma$ (5$\sigma$) sensitivities.  The sensitivity is significantly improved in a broad mass range.


\subsection{Case C:  $\tau$ final states}

\label{subsec:ditau}
For Case C, the signal channel at the LHC is $p p \ra Z^\prime + X$ with $Z^\prime$ decaying  to a tau pair. For a high-mass mediator decaying to di-tau, ATLAS~\cite{Aaboud:2017sjh} and CMS~\cite{Khachatryan:2016qkc} have set a $2\sigma$ upper limit on inclusive $\sigma(pp\ra Z^\prime + X ) \cdot B(Z^\prime\ra \tau^+\tau^-) $ in the 200~${\rm GeV} \lesssim M_{Z^\prime} \lesssim 4$~TeV (ATLAS) and 500~${\rm GeV}\lesssim M_{Z^\prime} \lesssim 3$~TeV (CMS) mass ranges with $\sqrt{s}=13$~TeV and 36.1~$\rm{fb}^{-1}$ and 2.2~$\rm{fb}^{-1}$, respectively. We only display the ATLAS constraint on $g^\prime$ in the lower panel of Fig.~\ref{fig:bounds}. 

We also estimate the sensitivity reach for 20~GeV~$\lesssim M_{Z^\prime} \lesssim 6000$ GeV at the HL-LHC with 3000 $\rm{fb}^{-1}$ of integrated luminosity.  There are mainly four decay modes for di-tau, namely, $\tau_e\tau_\mu$(6\%), $\tau_e\tau_h$(23\%), $\tau_\mu\tau_h$(23\%),  and $\tau_h\tau_h$(42\%), where $h$ denotes a hadron. In this analysis, we use the TauDecay package~\cite{Hagiwara:2012vz} to model the relatively clean leptonic and semi-leptonic decay modes of the taus. The main backgrounds for $\tau_e\tau_\mu$ are $t\bar{t}$, $WW$, and DY. For the semi-leptonic modes, the main backgrounds are DY and $W$+jets. To include the QCD multijet background in the semi-leptonic modes, we add 6\% and 28\% of the sum of the DY and $W$+jets backgrounds for the  $\tau_\mu\tau_h$ and $\tau_e\tau_h$ modes, respectively~\cite{Khachatryan:2016qkc}. The signal and DY background events are generated at LO and scaled by a K-factor of 1.3~\cite{Khachatryan:2016zqb} for $M_{\tau\tau} > M_Z$, while for $M_{\tau\tau} < M_Z$, we generate the signal and DY background with up to two additional jets in the final states.  We generate $t\bar{t}$, $WW$, and $W$+jets background events at LO. To take higher-order corrections into account, the LO cross section of $t\bar{t}$ is normalized to the NNLO + NNLL cross section by a factor of 1.84~\cite{web}. The LO cross sections of $WW$ and $W$+jets are normalized to NNLO QCD by a factor of 1.98~\cite{Gehrmann:2014fva} and 1.46~\cite{Boughezal:2016dtm}, respectively. To reduce the background, we implement two different selection rules SR1 and SR2 for $M_{Z^\prime}$ below and above the $Z$-pole. In the $\tau_e\tau_\mu$ mode, both SR1 and SR2 require:
	\beq
	\begin{split}
	&\bullet\; \text{Only one muon and one oppositely charged electron with } p_T > 20\; \text{GeV and}\ |\eta| < 2.4 ,\\
	&\bullet\;  \text{veto $b$-tagged jets},\\
	&\bullet\;   0.2 M_{Z^\prime} < M_{\tau_1\tau_2} < 0.8 M_{Z^\prime},\\
	&\bullet\;  M^{\mu}_T < 40\,\rm{GeV},\nonumber
	\end{split}
	\eeq
	where $\tau_1$ and $\tau_2$ are respectively $e$ and $\mu$, and $M^\mu_T$ is the transverse mass of the charged lepton $\mu$ and the missing transverse momentum $\vec{\slashed{E}}_T$ is defined as 
	$$M^\mu_T=\sqrt{2P_T^\mu \cdot \slashed{E}_T (1-\cos\Delta\phi(\mu,\vec{\slashed{E}}_T)) }\,.$$ 
	In addition, SR1 requires
	\beq
	\bullet\;  \Delta R(\tau_1,\tau_2) < \Delta R_{\rm{cut}}\,,
	\eeq
	where $\Delta R$ is the angular distance between $\tau_1$ and $\tau_2$. $\Delta R_{\rm{cut}}$ is varied with $M_{Z^\prime}$ to maximize the local significance $S_l$. For example, we choose $\Delta R_{\rm{cut}}$ = 1.0 (1.6) for $M_{Z^\prime} = 20\;(40)$ GeV.
	
	SR2 further requires
	\begin{eqnarray}
	\bullet\;  &&\cos \Delta\phi(\tau_1,\tau_2) < -0.95\,,\quad \\
	\bullet\;  &&\cos \Delta\phi(\tau_1,\vec{\slashed{E}}_T) + \cos \Delta\phi(\tau_2,\vec{\slashed{E}}_T) > -0.1\,,\quad \\
	\bullet\; &&\slashed{E}_T > \slashed{E}^{\rm{cut}}_T\,,
	\end{eqnarray}
	where the missing energy cut $\slashed{E}^{\rm{cut}}_T$ is varied with $M_{Z^\prime}$ to maximize the local significance $S_l$. We take $\slashed{E}^{\rm{cut}}_T$ to be 40 (450) GeV for $M_{Z^\prime} = 500\;(2000)$ GeV. 
	
	In the $\tau_\ell\tau_h$ modes, both SR1 and SR2 require:
	\beq
	\begin{split}
	\bullet\; &\text{Only one charged lepton and at least one opposite-sign tau-tagged jet with}\\
	&\;p_T > 20\; \text{GeV}\; \text{and } |\eta| < 2.4 ,\\
	\bullet\; &\text{veto $b$-tagged jets},\\
	\bullet\;  & 0.3 M_{Z^\prime} < M_{\tau_1\tau_2} < 0.9 M_{Z^\prime},\\
	\bullet\;  &M^{\ell}_T < 40\,\rm{GeV}.
	\end{split}
	\eeq
	The further requirements of SR1 and SR2 are the same as for the leptonic $\tau_e\tau_\mu$ mode, with $\tau_1$ and $\tau_2$ the charged lepton and tau-tagged jet, respectively. The blue solid (dashed) curve  in the lower panel of Fig.~\ref{fig:bounds} shows the $2\sigma$ ($5\sigma$) sensitivity for Case C using a combination of the three decay modes ($\tau_e\tau_\mu$, $\tau_e\tau_h$, and $\tau_\mu\tau_h$), respectively, with 3000 $\rm{fb}^{-1}$ at the HL-LHC. 

\section{Correlated signatures at CE$\nu$NS and collider experiments}
\label{sec:correlation}

It is of fundamental importance that we observe correlated signals of NSI in different experiments. 
In this section, we study correlated signatures at future CE$\nu$NS and collider experiments. 
We first simulate spectra in the presence of NSI and then examine the consistency between the two experiments in the hope of identifying a correlated signal. 
We select the benchmark point,
$$M_{Z^\prime}=10\ {\rm GeV\ and}\ \  g^\prime=0.002\,,$$ 
for Cases A and B and explore how a signal observed in one experiment will manifest in another. The point is marked with a star 
in Fig.~\ref{fig:bounds}. The point is chosen so that observable signals can be produced at COHERENT and at the LHC. Since this set of parameters does not produce a signal at DUNE and T2HK, we focus on correlated signatures at COHERENT with an upgraded LAr detector and the high luminosity LHC with $L = 3000\; \rm{fb}^{-1}$. 
Note that the benchmark point is chosen in a currently allowed narrow region near $m(\Upsilon(1S))$, and 
that LHCb  data impose strong constraints for $M_{Z^\prime}$ below and above it. 

\begin{figure}[t!]
	\begin{subfigure}{.5\textwidth}
		\centering
		\includegraphics[width=\textwidth]{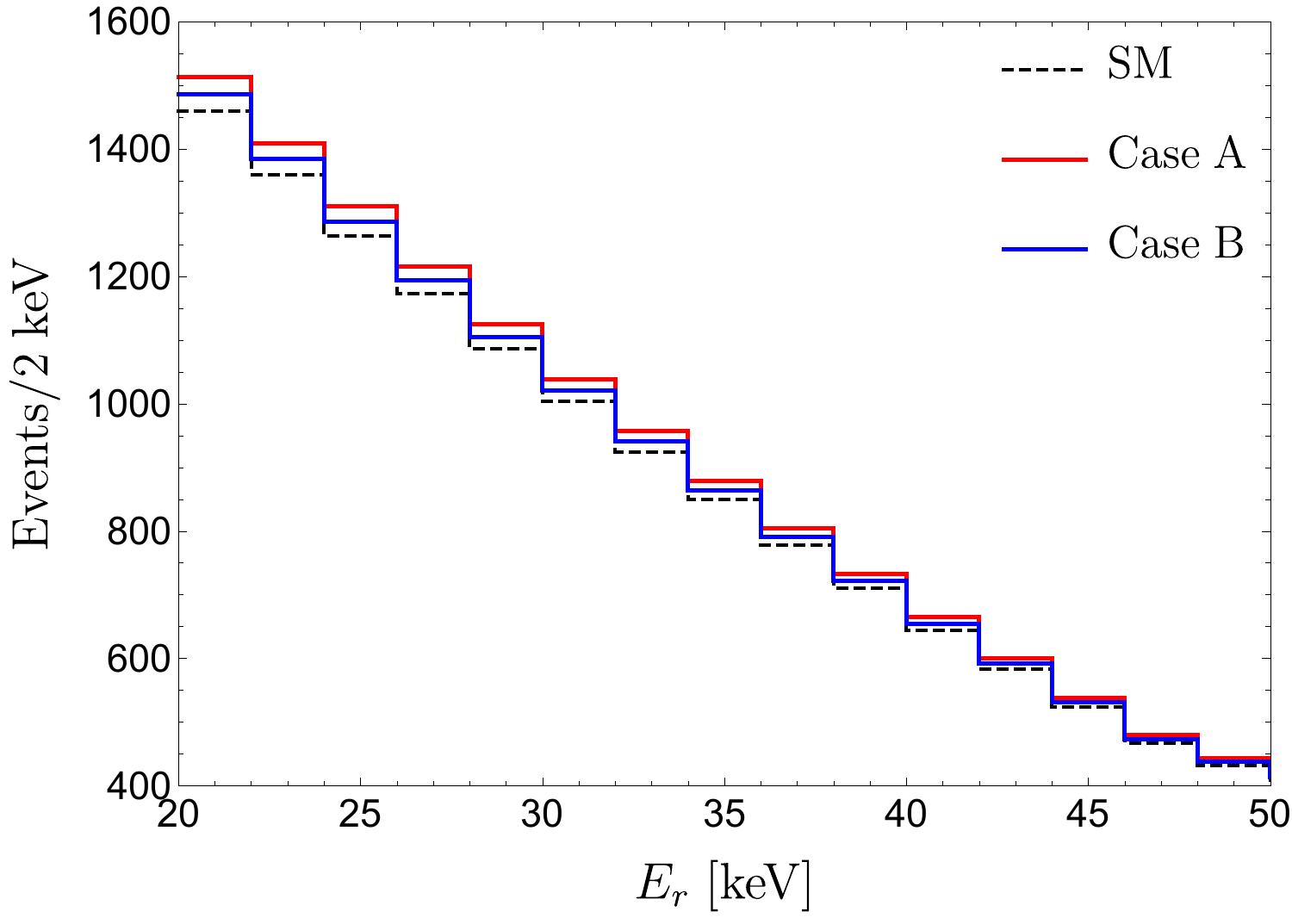}
		\label{fig:CEvNSenergy}
	\end{subfigure}
	\begin{subfigure}{.5\textwidth}
		\centering
		\includegraphics[width=\textwidth]{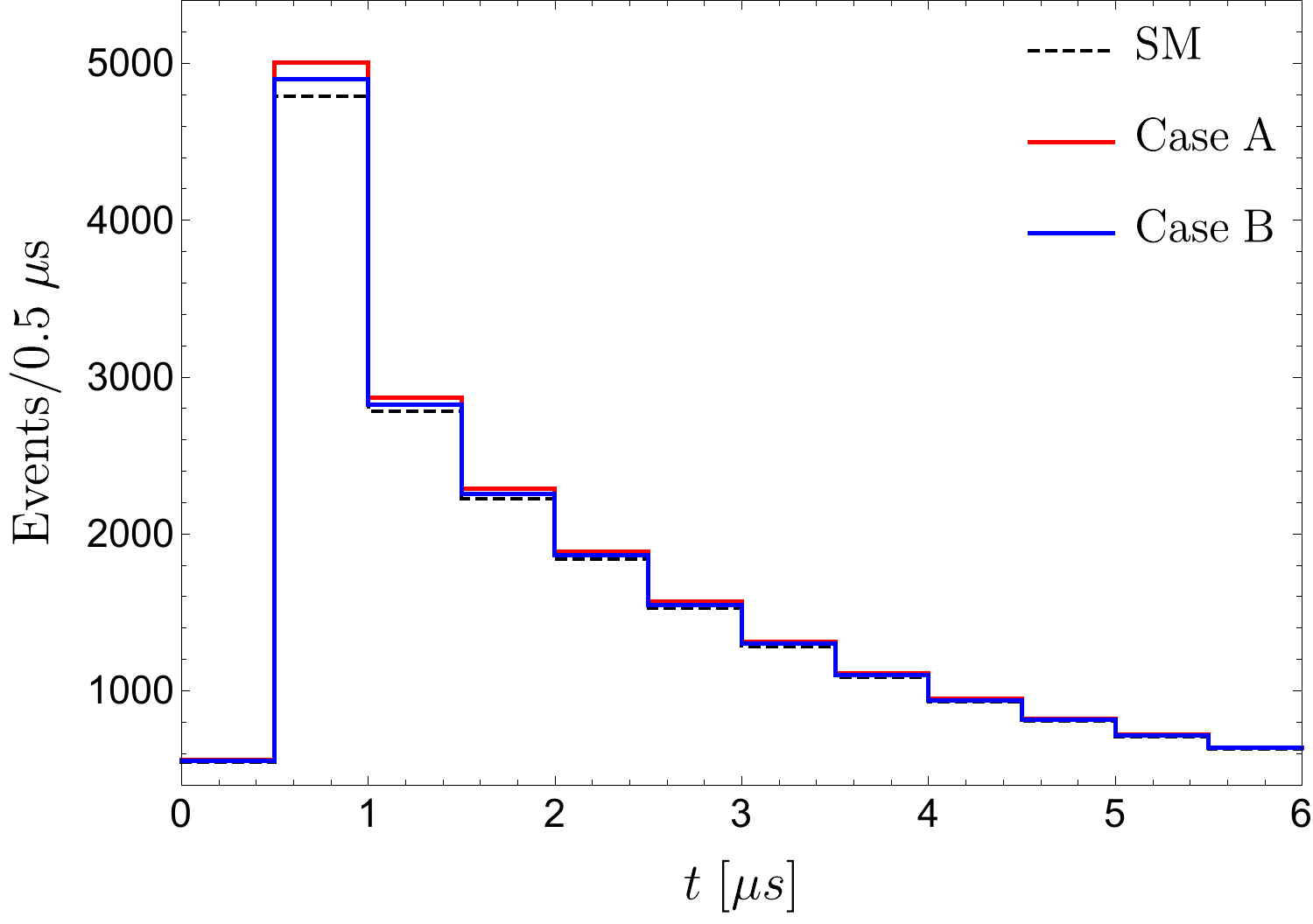}
		\label{fig:CEvNStime}
	\end{subfigure}
	\caption{Recoil energy (left) and temporal (right) distributions in an upgraded COHERENT LAr detector with $m_\text{det}=750$ kg and 
	4~years of data. The black dashed histograms correspond to the SM case, the red (blue) lines correspond to Case A (B) with $M_{Z^\prime}=10$ GeV and $g^{\prime}=0.002$.}
	\label{fig:CEvNS}
\end{figure}
\begin{figure}[t]
 	\begin{subfigure}{.5\textwidth}
 		\centering
 		\includegraphics[width=\textwidth]{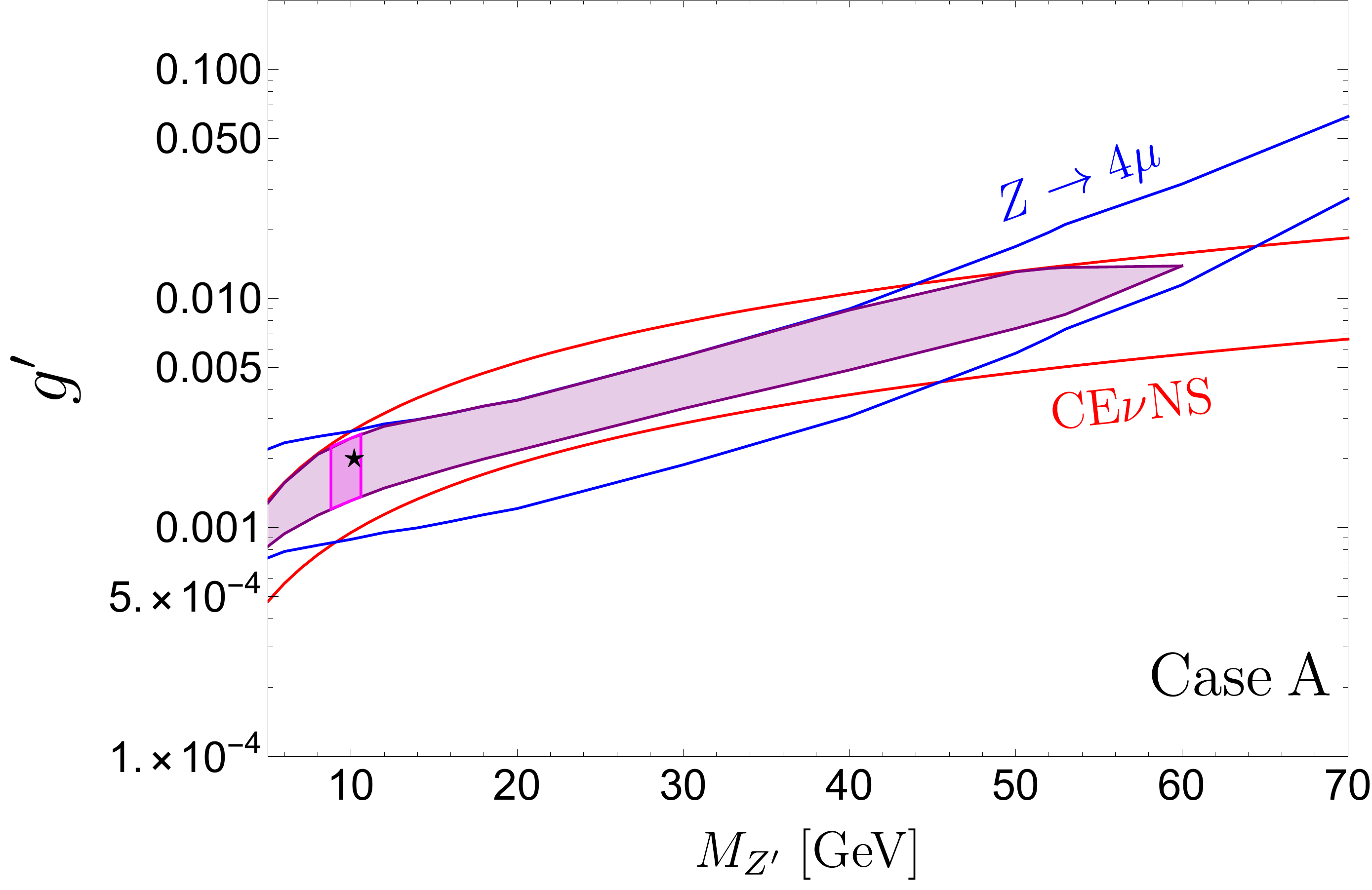}
 		\label{fig:counterA}
 	\end{subfigure}
 	\begin{subfigure}{.5\textwidth}
 		\centering
 		\includegraphics[width=\textwidth]{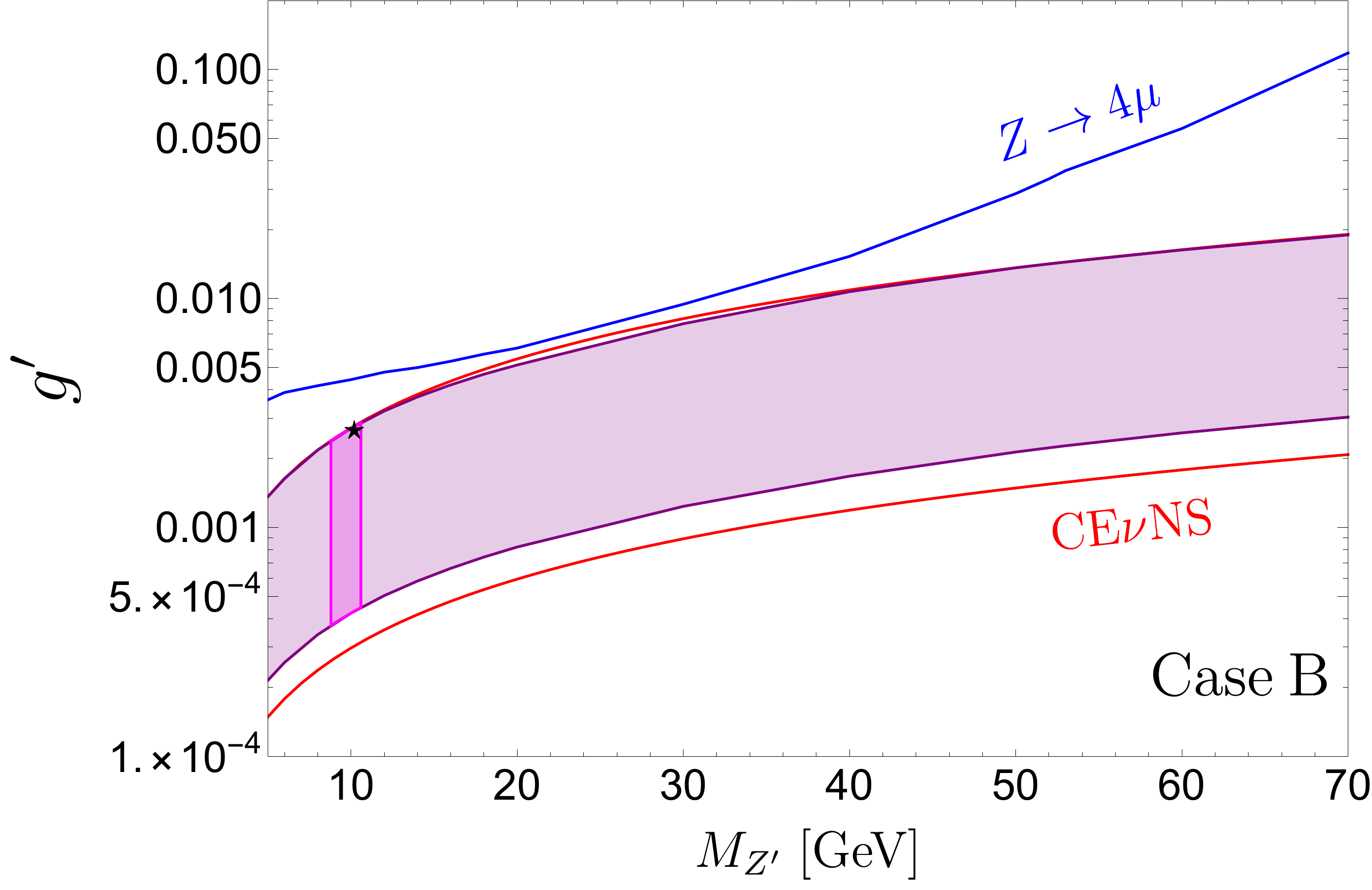}
 		\label{fig:counterB}
 	\end{subfigure}
 	\caption{$2\sigma$ allowed regions for Case A (left) and $1\sigma$ allowed regions for Case B (right) from COHERENT with a large LAr detector (within the red curves) and HL-LHC $Z \rightarrow 4 \mu$ decays (within the blue curves). The purple shaded regions ($2\sigma$ for case A and $1\sigma$ for Case B) are from our joint analysis. The magenta shaded regions are the allowed regions after including the LHCb bound as a prior. The stars mark the best fit points from our joint analysis.}
 	\label{fig:counter}
 \end{figure}

We first study signatures at COHERENT with an upgraded LAr detector. The recoil energy and temporal distributions of the events are shown in the left and right panel of Fig.~\ref{fig:CEvNS}, respectively. 
As can be seen from the left panel, the event excess is mainly at low energies. From the right panel, we see that the event excess peaks at around $t=1$~$\mu s$. This is due to the fact that the prompt component of the COHERENT flux is primarily composed of $\nu_\mu$, and the NSI coupling to $\nu_\mu$ leads to a modification of the number of events in Cases A and B. To analyze the spectra and to facilitate a joint analysis with simulated LHC data, we define
\beq
\chi^2(\vec{\theta}) = -2\; \text{ln} (L(\vec{\theta}))\,,
\label{eq:chi2_cevns}
\eeq
where $L(\vec{\theta})$ is defined in Eq.~(\ref{eq:L}) with $\vec{\theta} = \{g^\prime , M_{Z^\prime}\}$. We then calculate 
$\Delta \chi^2= \chi^2 - \chi^2_{\rm{min}}$.
The $2\sigma$ allowed region for Case A and $1\sigma$ allowed region for Case B, with data simulated with our benchmark point, are the regions between the red curves in Fig.~\ref{fig:counter}. The $2\sigma$ regions for Case B are too large to display.

\begin{figure}[t!]
 	\centering
        \includegraphics[width=0.45\textwidth]{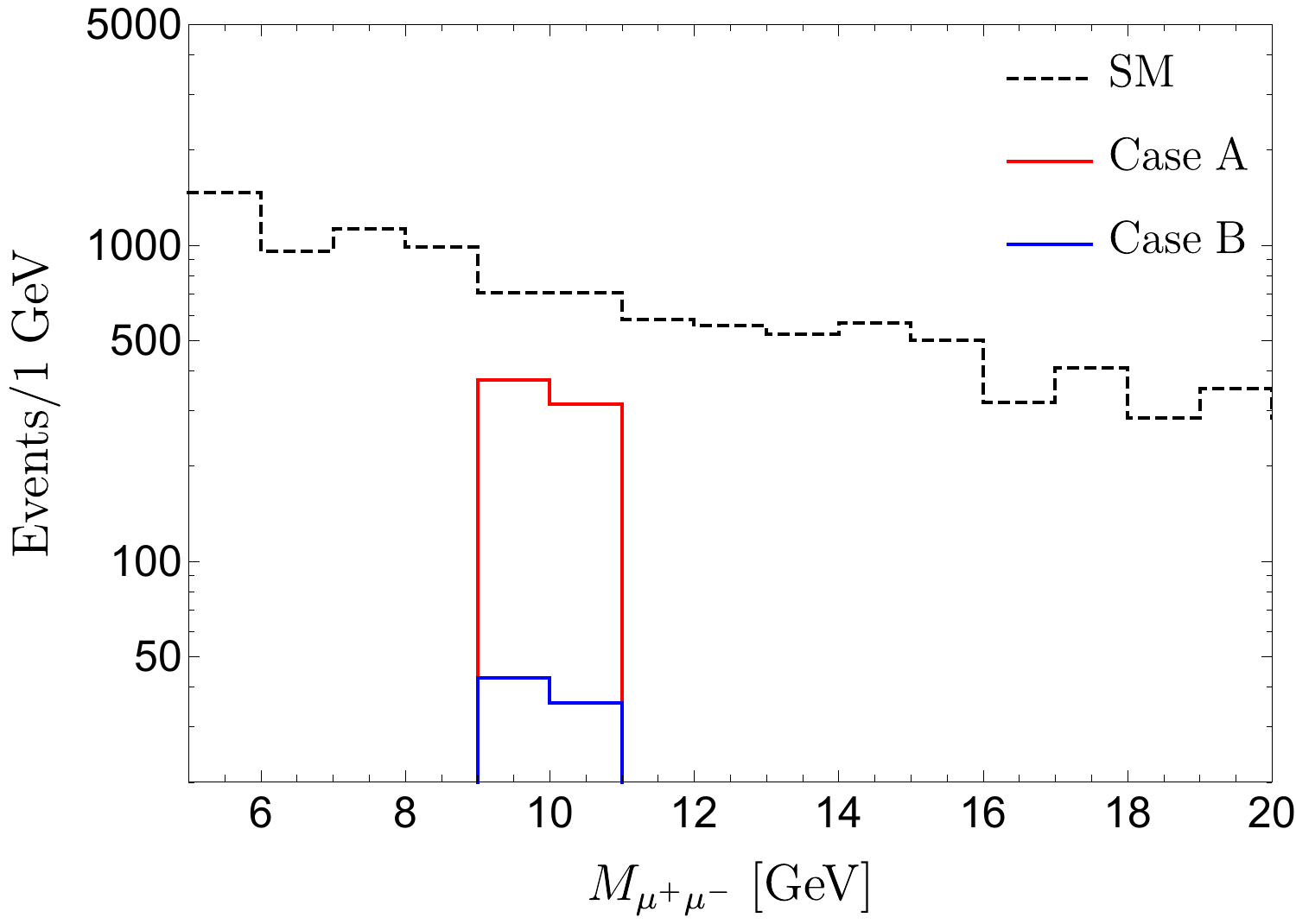}
 	\includegraphics[width=0.45\textwidth]{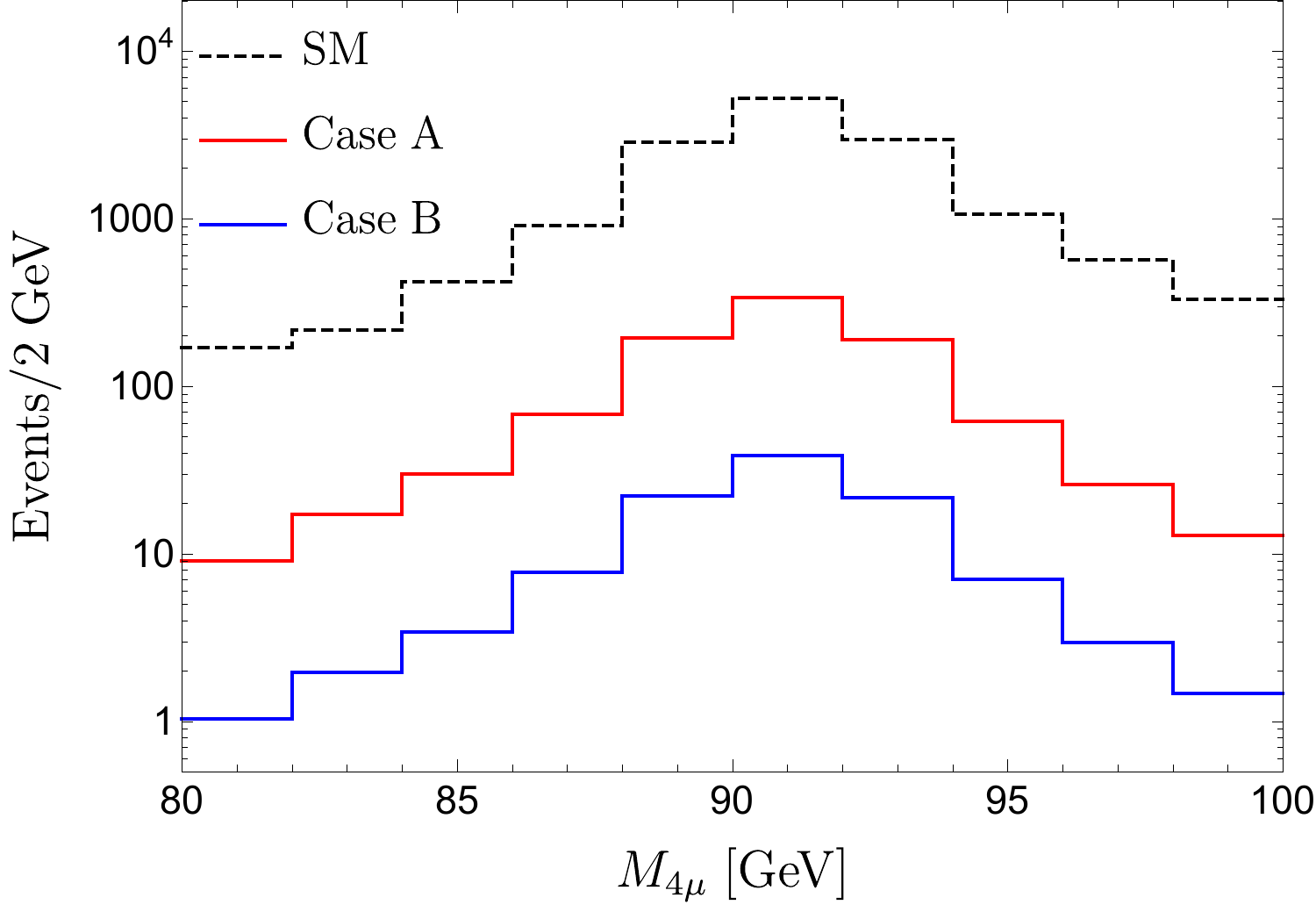}
 	\caption{Distributions of the reconstructed $M_{Z^\prime}$ (left) and  $M_{4\mu}$ (right) at the HL-LHC with $\sqrt{s} = 14$~TeV and $L=3000~\text{fb}^{-1}$ for $M_{Z^\prime}=10$~GeV and $g^{\prime}=0.002$, for Case A (red curves) and Case B (blue curves).}
 	\label{fig:4mu}
 \end{figure}

We now study signatures at the HL-LHC. Since we are interested in the low-mass region, we focus on the clean channel, $Z\rightarrow \mu^+\mu^- Z' \to 4 \mu$. We generate the leading process $q \bar{q}\rightarrow 4 \mu$ at the leading order (LO).
Following the CMS analysis~\cite{Sirunyan:2018nnz}, we require at least four well-identified and isolated muons to have $p_T > 5$ GeV and to be in the central region of the detector $|\eta|< 2.4$, with at least two muons to have $p_T > 10$ GeV and at least one to have $p_T > 20$ GeV. Dimuon candidates formed from an opposite sign muon pair are required to have $4 < M_{\mu^+\mu^-} < 120$ GeV. The four selected muons are required to have zero net charge and $80 < M_{4\mu} < 100$~GeV. The NNLO/LO K-factor is chosen to be 1.29~\cite{Sirunyan:2018nnz}. By following the CMS procedure in Ref.~\cite{Sirunyan:2018nnz}, we are able to reconstruct $M_{Z^\prime}$, whose distributions are shown in the left panel of Fig.~\ref{fig:4mu}. 
Unfortunately for $Z'$s of GeV mass, COHERENT sees an overall suppression in the CE$\nu$NS event rate, but no spectral distortion, thereby
precluding it from determining $M_{Z'}$. So a di-muon invariant mass cut cannot be applied and the look-elsewhere effect must be taken into account. Instead, we employ the $M_{4\mu}$ distributions (shown in the right panel of Fig.~\ref{fig:4mu}) to evaluate the precision with which
the $Z'$ parameters can be determined. We divide the range of $M_{4\mu}$ (80 GeV, 100 GeV) equally into 10~bins and perform a $\chi^2$ analysis with
 \beq
 \chi^2 = \sum_{i} \frac{N_{S,i}^2}{N_{B,i}+(\sigma_B N_{B,i})^2 }\,,\\
 \label{eq:chi2_4m}
 \eeq
  where $N_{S,i}$ ($N_{B,i}$) is the expected number of signal (background) events in the $i^{\rm{th}}$ bin. The background systematic uncertainty $\sigma_B$ is chosen to be 5\%. The parameters favored at $2\sigma$ for Case A and at $1\sigma $ for Case B lie between the blue curves in Fig.~\ref{fig:counter}; Case B has no lower blue curve because the SM is allowed at $1\sigma$. (The brown
  dashed curves in Fig.~\ref{fig:bounds} for the $2\sigma$ sensitivity to the $4\mu$ channel are produced by requiring the di-muon invariant mass $M_{\mu^+\mu^-}$ to be within 2\% of  $M_{Z'}$, and defining the local significance as
$N_S/\sqrt{N_B + \sigma_B^2 N_B^2}$.)
 
 We perform a joint  analysis of  future COHERENT and HL-LHC data by combining the two $\chi^2$ in Eqs.~(\ref{eq:chi2_cevns}) and~(\ref{eq:chi2_4m}). The resulting $2\sigma$ allowed regions for Case A and $1\sigma$ allowed regions for Case B are shaded in purple in Fig.~\ref{fig:counter}.  Consider Case A. The fact that the allowed regions from COHERENT and LHC have different slopes enables a combination of their datasets to limit $M_{Z'}$ to be below about 60~GeV. However, a precise determination of $M_{Z'}$ is not achieved even by combining the datasets. 
 For Case B, both COHERENT and HL-LHC only provide upper bounds on $g'$ at $2\sigma$.
 COHERENT dominates the sensitivity and the HL-LHC does not lead to a clear signal observation in the parameter region considered.
 
We now impose the stringent bounds from LHCb. To include the LHCb constraint, for each value of $M_{Z'}$ we add $\chi^2_{\rm{LHCb}} = 2.71 (g^\prime/g^\prime_{\rm{bound}})^2$ to our joint $\chi^2$, where $g^\prime_{\rm{bound}}$ is the 90\% CL exclusion limit from LHCb at that value of $M_{Z'}$; note that the LHCb dark photon search~\cite{Aaij:2019bvg} is performed independently at each mass, so that only one parameter, $g'$,
is varied in the analysis.
  On including the LHCb constraint, the allowed regions shrink significantly; see the magenta shaded regions in Fig.~\ref{fig:counter}. 

\section{Summary}
\label{sec:sum}

Next generation neutrino oscillation and CE$\nu$NS experiments will reach the sensitivity to discover new physics parameterized in the form of NSI. It is natural to seek complementary probes for NSI. Indeed, most beyond the Standard Model scenarios that generate NSI often result in simultaneous couplings to charged leptons, which opens up new possibilities to search for new physics associated with NSI at colliders.

In this work we studied a simple anomaly-free, ultraviolet-complete, gauged $U(1)'$ model that generates lepton flavor universality violating NSI. We considered three scenarios: $B-3L_\mu$, $B-\frac{3}{2}(L_\mu+L_\tau)$, and $B-3L_\tau$. The $Z'$ decay branching fractions are shown in Fig.~\ref{fig:decay}. Our main results are shown in Fig.~\ref{fig:bounds} and we summarize them as follows. 

\noindent
For constraints from current data:
\begin{enumerate}

	\item In Cases A and B, we mainly use neutrino oscillation, CE$\nu$NS, and collider experiments to put constraints on the coupling $g'$ in the mass range, 5~MeV$< M_{Z'}< 6$~TeV. We found that neutrino oscillation and CE$\nu$NS experiments give the most stringent bounds for masses below the dimuon threshold which is around 200~MeV. Above the dimuon threshold up to 70~GeV, LHCb prompt-like dark photon searches provide the strongest constraints except near the $J/\psi,\ \Upsilon$ resonances and in the vicinity of the $Z$-pole. ATLAS dimuon searches give the strongest bounds in the mass range,
	 $250\;{\rm{GeV}}\leq M_{Z'} \leq 6$ TeV. 
	
	\item The $(g-2)_\mu$ favored region is excluded by a combination of the experiments in the mass range considered.
	
	\item  Our Case C is unconstrained by the COHERENT experiment. Neutrino oscillation experiments set the strongest constraints up to 
	200~GeV. The LHC gives the strongest constraints for  $200\;{\rm{GeV}}\leq M_{Z'} \leq 4$~TeV. 

\end{enumerate}

\noindent
Our future projections are:
\begin{enumerate}

	\item We estimated the sensitivity of the high luminosity LHC with an integrated luminosity of 3~ab$^{-1}$ and find the that the reach of the $Z' \rightarrow \ell^+\ell^-$ channel is significantly improved in all of three scenarios; see Fig.~\ref{fig:bounds}.
	
	\item If the new gauge boson couples to  first and second generation leptons, future CE$\nu$NS data can set stronger bounds than next-generation neutrino oscillation experiments in almost the entire mass range.
	
	\item DUNE and T2HK have the best sensitivity for $Z'$ masses between $5-20$~MeV and $5-10$~MeV for Cases A and B, respectively.
	
	\item In Cases A and B, for $M_{Z'}$ above 10~GeV and couplings close to the sensitivity of the upgraded COHERENT experiment, in addition to  
	the CE$\nu$NS event numbers being modified, significant distortions in the four-muon invariant mass distribution in the $Z \rightarrow 4 \mu$ 
	search at the HL-LHC are expected.
	
	\item Combining CE$\nu$NS and collider data will help to limit $M_{Z'}$ from above; see Fig.~\ref{fig:counter}. 
\end{enumerate}

\acknowledgments
We thank Ahmed Ismail, Ernest Ma, Richard Ruiz and Kate Scholberg for helpful discussions. The work of T.H. and H.L. was supported in part by the U.S.~Department of Energy under grant No.~DE-FG02- 95ER40896 and in part by the PITT PACC. The work of J.L. was supported by the National Natural Science Foundation of China under Grant No.~11905299. D.M. was supported in part by the U.S. Department of Energy
under Grant No.~de-sc0010504. D.M. thanks PITT PACC, the Mainz Institute for Theoretical
Physics, the Tsung-Dao Lee Institute, Sun-Yat Sen University, and the Aspen Center for
Physics (which is supported by NSF Grant No. PHY-1607611) for their hospitality while this work was in progress.

\newpage
\bibliographystyle{JHEP}
\bibliography{ref}

\providecommand{\href}[2]{#2}\begingroup\raggedright\begin{thebibliography}{10}

\bibitem{Tanabashi:2018oca}
{\bf Particle Data Group} Collaboration, M.~Tanabashi et~al., {\it {Review of
  Particle Physics}},  {\em Phys. Rev.} {\bf D98} (2018), no.~3 030001.

\bibitem{Wolfenstein:1977ue}
L.~Wolfenstein, {\it {Neutrino Oscillations in Matter}},  {\em Phys. Rev.} {\bf
  D17} (1978) 2369--2374. [,294(1977)].

\bibitem{Ohlsson:2012kf}
T.~Ohlsson, {\it {Status of non-standard neutrino interactions}},  {\em Rept.
  Prog. Phys.} {\bf 76} (2013) 044201,
  [\href{http://arxiv.org/abs/1209.2710}{{\tt arXiv:1209.2710}}].

\bibitem{Miranda:2015dra}
O.~G. Miranda and H.~Nunokawa, {\it {Non standard neutrino interactions:
  current status and future prospects}},  {\em New J. Phys.} {\bf 17} (2015),
  no.~9 095002, [\href{http://arxiv.org/abs/1505.06254}{{\tt
  arXiv:1505.06254}}].

\bibitem{Farzan:2017xzy}
Y.~Farzan and M.~Tortola, {\it {Neutrino oscillations and Non-Standard
  Interactions}},  {\em Front.in Phys.} {\bf 6} (2018) 10,
  [\href{http://arxiv.org/abs/1710.09360}{{\tt arXiv:1710.09360}}].

\bibitem{Biggio:2009nt}
C.~Biggio, M.~Blennow, and E.~Fernandez-Martinez, {\it {General bounds on
  non-standard neutrino interactions}},  {\em JHEP} {\bf 08} (2009) 090,
  [\href{http://arxiv.org/abs/0907.0097}{{\tt arXiv:0907.0097}}].

\bibitem{Guzzo:1991hi}
M.~M. Guzzo, A.~Masiero, and S.~T. Petcov, {\it {On the MSW effect with
  massless neutrinos and no mixing in the vacuum}},  {\em Phys. Lett.} {\bf
  B260} (1991) 154--160. [,369(1991)].

\bibitem{Mikheev:1986gs}
S.~P. Mikheyev and A.~{\relax Yu}. Smirnov, {\it {Resonance Amplification of
  Oscillations in Matter and Spectroscopy of Solar Neutrinos}},  {\em Sov. J.
  Nucl. Phys.} {\bf 42} (1985) 913--917. [,305(1986)].

\bibitem{Akimov:2017ade}
{\bf COHERENT} Collaboration, D.~Akimov et~al., {\it {Observation of Coherent
  Elastic Neutrino-Nucleus Scattering}},  {\em Science} {\bf 357} (2017),
  no.~6356 1123--1126, [\href{http://arxiv.org/abs/1708.01294}{{\tt
  arXiv:1708.01294}}].

\bibitem{Freedman:1973yd}
D.~Z. Freedman, {\it {Coherent Neutrino Nucleus Scattering as a Probe of the
  Weak Neutral Current}},  {\em Phys. Rev.} {\bf D9} (1974) 1389--1392.

\bibitem{Barranco:2005yy}
J.~Barranco, O.~G. Miranda, and T.~I. Rashba, {\it {Probing new physics with
  coherent neutrino scattering off nuclei}},  {\em JHEP} {\bf 12} (2005) 021,
  [\href{http://arxiv.org/abs/hep-ph/0508299}{{\tt hep-ph/0508299}}].

\bibitem{Scholberg:2005qs}
K.~Scholberg, {\it {Prospects for measuring coherent neutrino-nucleus elastic
  scattering at a stopped-pion neutrino source}},  {\em Phys. Rev.} {\bf D73}
  (2006) 033005, [\href{http://arxiv.org/abs/hep-ex/0511042}{{\tt
  hep-ex/0511042}}].

\bibitem{Antusch:2008tz}
S.~Antusch, J.~P. Baumann, and E.~Fernandez-Martinez, {\it {Non-Standard
  Neutrino Interactions with Matter from Physics Beyond the Standard Model}},
  {\em Nucl. Phys.} {\bf B810} (2009) 369--388,
  [\href{http://arxiv.org/abs/0807.1003}{{\tt arXiv:0807.1003}}].

\bibitem{Kownacki:2016pmx}
C.~Kownacki, E.~Ma, N.~Pollard, and M.~Zakeri, {\it {Generalized Gauge U(1)
  Family Symmetry for Quarks and Leptons}},  {\em Phys. Lett.} {\bf B766}
  (2017) 149--152, [\href{http://arxiv.org/abs/1611.05017}{{\tt
  arXiv:1611.05017}}].

\bibitem{Konaka:1986cb}
A.~Konaka et~al., {\it {Search for Neutral Particles in Electron Beam Dump
  Experiment}},  {\em Phys. Rev. Lett.} {\bf 57} (1986) 659.

\bibitem{Riordan:1987aw}
E.~M. Riordan et~al., {\it {A Search for Short Lived Axions in an Electron Beam
  Dump Experiment}},  {\em Phys. Rev. Lett.} {\bf 59} (1987) 755.

\bibitem{Bjorken:1988as}
J.~D. Bjorken, S.~Ecklund, W.~R. Nelson, A.~Abashian, C.~Church, B.~Lu, L.~W.
  Mo, T.~A. Nunamaker, and P.~Rassmann, {\it {Search for Neutral Metastable
  Penetrating Particles Produced in the SLAC Beam Dump}},  {\em Phys. Rev.}
  {\bf D38} (1988) 3375.

\bibitem{Bross:1989mp}
A.~Bross, M.~Crisler, S.~H. Pordes, J.~Volk, S.~Errede, and J.~Wrbanek, {\it {A
  Search for Shortlived Particles Produced in an Electron Beam Dump}},  {\em
  Phys. Rev. Lett.} {\bf 67} (1991) 2942--2945.

\bibitem{Davier:1989wz}
M.~Davier and H.~Nguyen~Ngoc, {\it {An Unambiguous Search for a Light Higgs
  Boson}},  {\em Phys. Lett.} {\bf B229} (1989) 150--155.

\bibitem{Banerjee:2018vgk}
{\bf NA64} Collaboration, D.~Banerjee et~al., {\it {Search for a Hypothetical
  16.7 MeV Gauge Boson and Dark Photons in the NA64 Experiment at CERN}},  {\em
  Phys. Rev. Lett.} {\bf 120} (2018), no.~23 231802,
  [\href{http://arxiv.org/abs/1803.07748}{{\tt arXiv:1803.07748}}].

\bibitem{Heeck:2018nzc}
J.~Heeck, M.~Lindner, W.~Rodejohann, and S.~Vogl, {\it {Non-Standard Neutrino
  Interactions and Neutral Gauge Bosons}},  {\em SciPost Phys.} {\bf 6} (2019),
  no.~3 038, [\href{http://arxiv.org/abs/1812.04067}{{\tt arXiv:1812.04067}}].

\bibitem{Davidson:2003ha}
S.~Davidson, C.~Pena-Garay, N.~Rius, and A.~Santamaria, {\it {Present and
  future bounds on nonstandard neutrino interactions}},  {\em JHEP} {\bf 03}
  (2003) 011, [\href{http://arxiv.org/abs/hep-ph/0302093}{{\tt
  hep-ph/0302093}}].

\bibitem{Ibarra:2004pe}
A.~Ibarra, E.~Masso, and J.~Redondo, {\it {Systematic approach to
  gauge-invariant relations between lepton flavor violating processes}},  {\em
  Nucl. Phys.} {\bf B715} (2005) 523--535,
  [\href{http://arxiv.org/abs/hep-ph/0410386}{{\tt hep-ph/0410386}}].

\bibitem{Esteban:2018ppq}
I.~Esteban, M.~C. Gonzalez-Garcia, M.~Maltoni, I.~Martinez-Soler, and
  J.~Salvado, {\it {Updated Constraints on Non-Standard Interactions from
  Global Analysis of Oscillation Data}},  {\em JHEP} {\bf 08} (2018) 180,
  [\href{http://arxiv.org/abs/1805.04530}{{\tt arXiv:1805.04530}}].

\bibitem{Akimov:2018ghi}
{\bf COHERENT} Collaboration, D.~Akimov et~al., {\it {COHERENT 2018 at the
  Spallation Neutron Source}},  \href{http://arxiv.org/abs/1803.09183}{{\tt
  arXiv:1803.09183}}.

\bibitem{Sirunyan:2018nnz}
{\bf CMS} Collaboration, A.~M. Sirunyan et~al., {\it {Search for an
  $L_{\mu}-L_{\tau}$ gauge boson using Z$\to4\mu$ events in proton-proton
  collisions at $\sqrt{s} =$ 13 TeV}},  {\em Phys. Lett.} {\bf B792} (2019)
  345--368, [\href{http://arxiv.org/abs/1808.03684}{{\tt arXiv:1808.03684}}].

\bibitem{TheBABAR:2016rlg}
{\bf BaBar} Collaboration, J.~P. Lees et~al., {\it {Search for a muonic dark
  force at BABAR}},  {\em Phys. Rev.} {\bf D94} (2016), no.~1 011102,
  [\href{http://arxiv.org/abs/1606.03501}{{\tt arXiv:1606.03501}}].

\bibitem{Aaij:2019bvg}
{\bf LHCb} Collaboration, R.~Aaij et~al., {\it {Search for
  $A'\!\to\!\mu^+\mu^-$ decays}},  \href{http://arxiv.org/abs/1910.06926}{{\tt
  arXiv:1910.06926}}.

\bibitem{Aad:2019fac}
{\bf ATLAS} Collaboration, G.~Aad et~al., {\it {Search for high-mass dilepton
  resonances using 139 fb$^{-1}$ of $pp$ collision data collected at
  $\sqrt{s}=$13 TeV with the ATLAS detector}},
  \href{http://arxiv.org/abs/1903.06248}{{\tt arXiv:1903.06248}}.

\bibitem{Aaboud:2017sjh}
{\bf ATLAS} Collaboration, M.~Aaboud et~al., {\it {Search for additional heavy
  neutral Higgs and gauge bosons in the ditau final state produced in 36
  fb$^{-1}$ of pp collisions at $ \sqrt{s}=13 $ TeV with the ATLAS detector}},
  {\em JHEP} {\bf 01} (2018) 055, [\href{http://arxiv.org/abs/1709.07242}{{\tt
  arXiv:1709.07242}}].

\bibitem{Mishra:1991bv}
{\bf CCFR} Collaboration, S.~R. Mishra et~al., {\it {Neutrino tridents and W Z
  interference}},  {\em Phys. Rev. Lett.} {\bf 66} (1991) 3117--3120.

\bibitem{Altmannshofer:2014pba}
W.~Altmannshofer, S.~Gori, M.~Pospelov, and I.~Yavin, {\it {Neutrino Trident
  Production: A Powerful Probe of New Physics with Neutrino Beams}},  {\em
  Phys. Rev. Lett.} {\bf 113} (2014) 091801,
  [\href{http://arxiv.org/abs/1406.2332}{{\tt arXiv:1406.2332}}].

\bibitem{Jegerlehner:2009ry}
F.~Jegerlehner and A.~Nyffeler, {\it {The Muon g-2}},  {\em Phys. Rept.} {\bf
  477} (2009) 1--110, [\href{http://arxiv.org/abs/0902.3360}{{\tt
  arXiv:0902.3360}}].

\bibitem{Liao:2016hsa}
J.~Liao, D.~Marfatia, and K.~Whisnant, {\it {Degeneracies in long-baseline
  neutrino experiments from nonstandard interactions}},  {\em Phys. Rev.} {\bf
  D93} (2016), no.~9 093016, [\href{http://arxiv.org/abs/1601.00927}{{\tt
  arXiv:1601.00927}}].

\bibitem{DUNE}
{\bf DUNE} Collaboration, R.~Acciarri et~al., {\it {Long-Baseline Neutrino
  Facility (LBNF) and Deep Underground Neutrino Experiment (DUNE)}},
  \href{http://arxiv.org/abs/1512.06148}{{\tt arXiv:1512.06148}}.

\bibitem{T2HK}
{\bf Hyper-Kamiokande} Collaboration, K.~Abe et~al., {\it {Hyper-Kamiokande
  Design Report}},  \href{http://arxiv.org/abs/1805.04163}{{\tt
  arXiv:1805.04163}}.

\bibitem{Liao:2016orc}
J.~Liao, D.~Marfatia, and K.~Whisnant, {\it {Nonstandard neutrino interactions
  at DUNE, T2HK and T2HKK}},  {\em JHEP} {\bf 01} (2017) 071,
  [\href{http://arxiv.org/abs/1612.01443}{{\tt arXiv:1612.01443}}].

\bibitem{SNS}
{\bf COHERENT} Collaboration, D.~Akimov et~al., {\it {The COHERENT Experiment
  at the Spallation Neutron Source}},
  \href{http://arxiv.org/abs/1509.08702}{{\tt arXiv:1509.08702}}.

\bibitem{Akimov:2018vzs}
{\bf COHERENT} Collaboration, D.~Akimov et~al., {\it {COHERENT Collaboration
  data release from the first observation of coherent elastic neutrino-nucleus
  scattering}},  \href{http://arxiv.org/abs/1804.09459}{{\tt
  arXiv:1804.09459}}.

\bibitem{Klein:1999gv}
S.~R. Klein and J.~Nystrand, {\it {Interference in exclusive vector meson
  production in heavy ion collisions}},  {\em Phys. Rev. Lett.} {\bf 84} (2000)
  2330--2333, [\href{http://arxiv.org/abs/hep-ph/9909237}{{\tt
  hep-ph/9909237}}].

\bibitem{Collar:2019ihs}
J.~I. Collar, A.~R.~L. Kavner, and C.~M. Lewis, {\it {Response of CsI[Na] to
  Nuclear Recoils: Impact on Coherent Elastic Neutrino-Nucleus Scattering
  (CE$\nu$NS)}},  \href{http://arxiv.org/abs/1907.04828}{{\tt
  arXiv:1907.04828}}.

\bibitem{Konovalov}
A.~Konovalov, {\it private communication}.

\bibitem{Dutta:2019eml}
B.~Dutta, S.~Liao, S.~Sinha, and L.~E. Strigari, {\it {Searching for Beyond the
  Standard Model Physics with COHERENT Energy and Timing Data}},
  \href{http://arxiv.org/abs/1903.10666}{{\tt arXiv:1903.10666}}.

\bibitem{AristizabalSierra:2019zmy}
D.~Aristizabal~Sierra, J.~Liao, and D.~Marfatia, {\it {Impact of form factor
  uncertainties on interpretations of coherent elastic neutrino-nucleus
  scattering data}},  {\em JHEP} {\bf 06} (2019) 141,
  [\href{http://arxiv.org/abs/1902.07398}{{\tt arXiv:1902.07398}}].

\bibitem{Alwall:2014hca}
J.~Alwall, R.~Frederix, S.~Frixione, V.~Hirschi, F.~Maltoni, O.~Mattelaer,
  H.~S. Shao, T.~Stelzer, P.~Torrielli, and M.~Zaro, {\it {The automated
  computation of tree-level and next-to-leading order differential cross
  sections, and their matching to parton shower simulations}},  {\em JHEP} {\bf
  07} (2014) 079, [\href{http://arxiv.org/abs/1405.0301}{{\tt
  arXiv:1405.0301}}].

\bibitem{Ball:2013hta}
{\bf NNPDF} Collaboration, R.~D. Ball, V.~Bertone, S.~Carrazza, L.~Del~Debbio,
  S.~Forte, A.~Guffanti, N.~P. Hartland, and J.~Rojo, {\it {Parton
  distributions with QED corrections}},  {\em Nucl. Phys.} {\bf B877} (2013)
  290--320, [\href{http://arxiv.org/abs/1308.0598}{{\tt arXiv:1308.0598}}].

\bibitem{Alloul:2013bka}
A.~Alloul, N.~D. Christensen, C.~Degrande, C.~Duhr, and B.~Fuks, {\it
  {FeynRules 2.0 - A complete toolbox for tree-level phenomenology}},  {\em
  Comput. Phys. Commun.} {\bf 185} (2014) 2250--2300,
  [\href{http://arxiv.org/abs/1310.1921}{{\tt arXiv:1310.1921}}].

\bibitem{Sjostrand:2006za}
T.~Sjostrand, S.~Mrenna, and P.~Z. Skands, {\it {PYTHIA 6.4 Physics and
  Manual}},  {\em JHEP} {\bf 05} (2006) 026,
  [\href{http://arxiv.org/abs/hep-ph/0603175}{{\tt hep-ph/0603175}}].

\bibitem{Sjostrand:2007gs}
T.~Sjostrand, S.~Mrenna, and P.~Z. Skands, {\it {A Brief Introduction to PYTHIA
  8.1}},  {\em Comput. Phys. Commun.} {\bf 178} (2008) 852--867,
  [\href{http://arxiv.org/abs/0710.3820}{{\tt arXiv:0710.3820}}].

\bibitem{Alwall:2007fs}
J.~Alwall et~al., {\it {Comparative study of various algorithms for the merging
  of parton showers and matrix elements in hadronic collisions}},  {\em Eur.
  Phys. J.} {\bf C53} (2008) 473--500,
  [\href{http://arxiv.org/abs/0706.2569}{{\tt arXiv:0706.2569}}].

\bibitem{deFavereau:2013fsa}
{\bf DELPHES 3} Collaboration, J.~de~Favereau, C.~Delaere, P.~Demin,
  A.~Giammanco, V.~Lematre, A.~Mertens, and M.~Selvaggi, {\it {DELPHES 3, A
  modular framework for fast simulation of a generic collider experiment}},
  {\em JHEP} {\bf 02} (2014) 057, [\href{http://arxiv.org/abs/1307.6346}{{\tt
  arXiv:1307.6346}}].

\bibitem{Khachatryan:2016zqb}
{\bf CMS} Collaboration, V.~Khachatryan et~al., {\it {Search for narrow
  resonances in dilepton mass spectra in proton-proton collisions at $\sqrt{s}$
  = 13 TeV and combination with 8 TeV data}},  {\em Phys. Lett.} {\bf B768}
  (2017) 57--80, [\href{http://arxiv.org/abs/1609.05391}{{\tt
  arXiv:1609.05391}}].

\bibitem{Ilten:2018crw}
P.~Ilten, Y.~Soreq, M.~Williams, and W.~Xue, {\it {Serendipity in dark photon
  searches}},  {\em JHEP} {\bf 06} (2018) 004,
  [\href{http://arxiv.org/abs/1801.04847}{{\tt arXiv:1801.04847}}].

\bibitem{Barze:2013fru}
L.~Barze, G.~Montagna, P.~Nason, O.~Nicrosini, F.~Piccinini, and A.~Vicini,
  {\it {Neutral current Drell-Yan with combined QCD and electroweak corrections
  in the POWHEG BOX}},  {\em Eur. Phys. J.} {\bf C73} (2013), no.~6 2474,
  [\href{http://arxiv.org/abs/1302.4606}{{\tt arXiv:1302.4606}}].

\bibitem{web}
https://twiki.cern.ch/twiki/bin/view/LHCPhysics/TtbarNNLO.

\bibitem{Kidonakis:2010ux}
N.~Kidonakis, {\it {Two-loop soft anomalous dimensions for single top quark
  associated production with a W- or H-}},  {\em Phys. Rev.} {\bf D82} (2010)
  054018, [\href{http://arxiv.org/abs/1005.4451}{{\tt arXiv:1005.4451}}].

\bibitem{Gehrmann:2014fva}
T.~Gehrmann, M.~Grazzini, S.~Kallweit, P.~Maierh{\~A}¶fer, A.~von Manteuffel,
  S.~Pozzorini, D.~Rathlev, and L.~Tancredi, {\it {$W^+W^-$ Production at
  Hadron Colliders in Next to Next to Leading Order QCD}},  {\em Phys. Rev.
  Lett.} {\bf 113} (2014), no.~21 212001,
  [\href{http://arxiv.org/abs/1408.5243}{{\tt arXiv:1408.5243}}].

\bibitem{Grazzini:2016swo}
M.~Grazzini, S.~Kallweit, D.~Rathlev, and M.~Wiesemann, {\it {$W^{\pm}Z$
  production at hadron colliders in NNLO QCD}},  {\em Phys. Lett.} {\bf B761}
  (2016) 179--183, [\href{http://arxiv.org/abs/1604.08576}{{\tt
  arXiv:1604.08576}}].

\bibitem{Cascioli:2014yka}
F.~Cascioli, T.~Gehrmann, M.~Grazzini, S.~Kallweit, P.~Maierh{\~A}¶fer, A.~von
  Manteuffel, S.~Pozzorini, D.~Rathlev, L.~Tancredi, and E.~Weihs, {\it {ZZ
  production at hadron colliders in NNLO QCD}},  {\em Phys. Lett.} {\bf B735}
  (2014) 311--313, [\href{http://arxiv.org/abs/1405.2219}{{\tt
  arXiv:1405.2219}}].

\bibitem{Khachatryan:2016qkc}
{\bf CMS} Collaboration, V.~Khachatryan et~al., {\it {Search for heavy
  resonances decaying to tau lepton pairs in proton-proton collisions at $
  \sqrt{s}=13 $ TeV}},  {\em JHEP} {\bf 02} (2017) 048,
  [\href{http://arxiv.org/abs/1611.06594}{{\tt arXiv:1611.06594}}].

\bibitem{Hagiwara:2012vz}
K.~Hagiwara, T.~Li, K.~Mawatari, and J.~Nakamura, {\it {TauDecay: a library to
  simulate polarized tau decays via FeynRules and MadGraph5}},  {\em Eur. Phys.
  J.} {\bf C73} (2013) 2489, [\href{http://arxiv.org/abs/1212.6247}{{\tt
  arXiv:1212.6247}}].

\bibitem{Boughezal:2016dtm}
R.~Boughezal, X.~Liu, and F.~Petriello, {\it {W-boson plus jet differential
  distributions at NNLO in QCD}},  {\em Phys. Rev.} {\bf D94} (2016), no.~11
  113009, [\href{http://arxiv.org/abs/1602.06965}{{\tt arXiv:1602.06965}}].

\end{thebibliography}\endgroup

\end{document}